\documentclass
[floatfix,superscriptaddress,secnumarabic,amssymb,amsmath,nobibnotes,aps,prd,showkeys,showpacs,nofootinbib,onecolumn,notitlepage,10pt]{revtex4}%
\usepackage{setspace}
\usepackage{color}
\usepackage{amsmath}
\usepackage{amsfonts}
\usepackage{verbatim}
\usepackage{amssymb}
\usepackage{graphicx,bm}
\usepackage{graphicx}
\usepackage{amsmath}
\usepackage{amssymb}
\usepackage{amssymb}
\usepackage{graphicx,bm}
\usepackage{graphicx}
\usepackage[caption=false]{subfig}%
\providecommand{\U}[1]{\protect\rule{.1in}{.1in}}

\newcommand{\be}{\begin{equation}}
\newcommand{\ee}{\end{equation}}

\newcommand{\mincir}{\raise
-3.truept\hbox{\rlap{\hbox{$\sim$}}\raise4.truept\hbox{$<$}\ }}
\newcommand{\magcir}{\raise
-3.truept\hbox{\rlap{\hbox{$\sim$}}\raise4.truept\hbox{$>$}\ }}

\ifx\pdfoutput\relax\let\pdfoutput=\undefined\fi
\newcount\msipdfoutput
\ifx\pdfoutput\undefined\else
\ifcase\pdfoutput\else
\ifx\paperwidth\undefined\else
\ifdim\paperheight=0pt\relax\else\pdfpageheight\paperheight\fi
\ifdim\paperwidth=0pt\relax\else\pdfpagewidth\paperwidth\fi
\fi\fi\fi
\begin{document}
\title{Observational Constraints on Noncoincident $f(Q)$-Gravity with Matter-Gravity Coupling}
\author{Andronikos Paliathanasis}
\email{anpaliat@phys.uoa.gr}
\affiliation{Institute of Systems Science, Durban University of Technology, Durban 4000,
South Africa}
\affiliation{Centre for Space Research, North-West University, Potchefstroom 2520, South Africa}
\affiliation{Departamento de Matem\`{a}ticas, Universidad Cat\`{o}lica del Norte, Avda.
Angamos 0610, Casilla 1280 Antofagasta, Chile}
\affiliation{National Institute for Theoretical and Computational Sciences (NITheCS), South Africa}

\begin{abstract}
We investigate $f\left(  Q\right)  $-gravity with a matter-gravity coupling as
a geometric dark energy candidate for the description of the late-time cosmic
acceleration within a spatially flat Friedmann--Lema\^{\i}tre-Robertson-Walker
geometry. We select a noncoincident connection that naturally follows from the
general framework of cosmological models with nonzero spatial curvature. We
present observational constraints for the simplest $f\left(  Q\right)
=f_{0}Q^{n}$ model using data from Supernovae, Baryon Acoustic Oscillations
and Cosmic Chronometers. For different data combinations we found consistent
constraints, with a best-fit value for the power-law index $n\simeq2$. A
comparison with the $\Lambda$CDM model shows that the $f\left(  Q\right)
$-gravity leads to larger values for the likelihood, while Akaike's
Information Criterion suggests statistical equivalence between the two models
for most data combinations.

\end{abstract}
\maketitle

\section{Introduction}

\label{sec1}

The study of the cosmological observations reveals that the universe is
currently undergoing accelerated expansion
\cite{SupernovaSearchTeam:1998fmf,SDSS:2003tbn,SupernovaCosmologyProject:2008ojh,Breton:2013twa}%
. The source of this cosmic acceleration remains unknown and is attributed to
dark energy \cite{Frieman:2008sn}. On the other hand, recent observations
reinforce earlier indications \cite{deCruzPerez:2024shj,Park:2024vrw} of a
dynamical dark energy component
\cite{DES:2025key,DESI:2025fii,DESI:2025zgx,DESI:2025zpo}, challenging the
dominance of the cosmological constant in describing the large scale structure
of the universe
\cite{Feng:2011zzo,Barboza:2008rh,Bernardo:2021cxi,Rezaei:2024vtg,Escamilla:2024fzq,Hur:2025lqc,Chakraborty:2025rvc,Luongo:2024zhc}%
.

In General Relativity, dark energy is the fluid component required to drive
the cosmic dynamics and account for the observed acceleration of the universe.
For this purpose, various proposals have been put forward in the literature
including scalar field models and many
others.\cite{Ratra:1987rm,Tsujikawa:2013fta,Gorini:2004by,vonMarttens:2022xyr,dePutter:2007ny,Bagla:2002yn,Farnes:2017gbf}%
. Nevertheless, in gravitational theories with modified Einstein-Hilbert
Action, cosmic acceleration follows naturally from the geometric
characteristics of the gravitational field, leading to a geometric description
of dark energy
\cite{Tsujikawa:2007xu,Nojiri:2017ncd,Bahamonde:2021gfp,Nojiri:2008nt,Heisenberg:2018vsk,Csillag:2025gnz,Chaudhary:2024jkj}%
.

A series of alternative gravitational theories that have recently drawn
attention in the literature share the common feature of selecting a flat and
symmetric affine connection, thus forming the family of extended symmetric
teleparallel theories
\cite{BeltranJimenez:2019tme,Khyllep:2021pcu,Zhao:2024kri,De:2023xua,Gakis:2019rdd,Nojiri:2024zab,Heisenberg:2023lru,Nojiri:2024hau,Carloni:2024ybx,Carloni:2025kev}%
. Within Symmetric Teleparallel General Relativity (STEGR)
\cite{Nester:1998mp}, gravity is described by a non-Riemannian manifold with a
symmetric and flat connection. The gravitational field is a result of the
nonmetricity scalar $Q$ \cite{Conroy:2017yln}. The connection is flat, which
means, there exists always a coordinate system known as the coincidence gauge
\cite{Zhao:2021zab}, where the covariant derivative is represented as partial
derivative. Hence, in this gravitational model, the inertia effects can be
separated from the gravitational field \cite{BeltranJimenez:2019tme}. As we
shall discuss in the following, the flat connection considered is essential
for the description of the gravitational field, and different connections lead
to different formulations for the gravitational field \cite{Hohmann:2021ast}.

The impact of the different connections in isotropic cosmological studies
within extended symmetric teleparallel theories has been investigated in
detail in \cite{Dimakis:2022rkd,Ayuso:2025vkc}, while for the case of
anisotropic cosmologies it has been examined in
\cite{Dimakis:2023uib,Murtaza:2025gme}. Furthermore, it was found that, in a
Kantowski--Sachs background, fluids with a tilted velocity can be supported
\cite{Paliathanasis:2024xpy}, in contrast with General Relativity, where this
feature is not possible for such a geometry. Moreover, the role of the
connection in astrophysical objects has been investigated in
\cite{Bahamonde:2022zgj,DAmbrosio:2021zpm}, with analytic solutions presented
in \cite{Dimakis:2024fan}.

The present work is focused on the extended $f\left(  Q\right)  $-theory
\cite{BeltranJimenez:2019tme}.~In this theoretic framework, the dynamics of
the gravitational field equations are determined by the nonlinear function $f$
together with the characteristics of the connection. Indeed, in the
coincidence gauge the connection makes no contribution to the gravitational
dynamics. By contrast, in a noncoincident choice, additional degrees of
freedom, effectively described by a scalar field, become dynamical and drive
the gravitational evolution. \cite{Paliathanasis:2023pqp}.

The global dynamics of $f\left(  Q\right)  $-gravity within a
Friedmann--Lema\^{\i}tre--Robertson--Walker (FLRW) geometry have been
investigated before in \cite{Paliathanasis:2023nkb,Paliathanasis:2023raj}. It
was found that, even for a common $f(Q)$ model, the evolution of the physical
quantities differs for each connection. For the theory with the coincidence
connection in order to describe the cosmic acceleration, it is necessary to
introduce a complex functional form for the $f\left(  Q\right)  $-theory, or
introduce the cosmological constant
\cite{Lazkoz:2019sjl,Anagnostopoulos:2021ydo,Atayde:2021pgb,Boiza:2025xpn,Ferreira:2023awf,Shi:2023kvu}%
. However, this is not true in the selection of a noncoincidence connection
where the nonlinear function $f$ model with the minimum number of free
parameters can provide dark energy dynamics \cite{Paliathanasis:2025hjw} and
explain the cosmic evolution. A similar correspondence between the connection
and the physical theory has been identified in other extended STEGR theories
\cite{Murtaza:2025klz,Murtaza:2025gme}.

Recently, in \cite{Abebe:2025wos} in the symmetric teleparallel $f\left(
Q\right)  $-gravity it was introduced a matter-gravity coupling function in
the Action Integral. It was found that the presence of the coupling function
allows for the existence of a matter-dominated era in the cosmological
history, which is otherwise absent in the case without matter-gravity
interaction. Moreover, the coupling function can be used to overpass the
pathologies that $f\left(  Q\right)  $-gravity suffers \cite{Gomes:2023tur},
such as strong coupling and the appearance of ghosts.

Cosmological scenarios with energy transfer between the elements of the dark
sector of the universe, that is, dark matter and dark energy, have attracted
interest recently, as they offer an alternative framework for describing the
observational data
\cite{Wang:2016lxa,Lucca:2021dxo,Figueruelo:2026eis,Yang:2018qec,Yang:2019vni,Pan:2020zza,Benisty:2024lmj,Aboubrahim:2026tks}%
. In interacting models, an effective phantom-like behavior of dark energy
does not correspond to a fundamental physical pathology, but instead emerges
as an apparent feature \cite{Guedezounme:2025wav}. Several works in the
literature indicate that the interacting scenario is supported by late-time
cosmological observations \cite{Zhai:2025hfi,Paliathanasis:2026ymi,Li:2026xaz}%
. Nevertheless, the majority of the interacting models discussed in the
literature are phenomenological.

In theoretical framework, interactions follows when a coupling function exists
in the gravitational Action Integral, providing nonzero interacting components
in the dark sector. Weyl Integrable Spacetime is a gravitational model that
provides, within a geometric framework, an interacting term
\cite{Salim:1996ei,Romero:2012hs}. The interaction follows as the result of
the selection of the connection for a conformally related metric. In
Scala-tensor theories, conformal transformations which connect the Jordan and
the Einstein frames \cite{Tsamparlis:2013aza} can be used to introduce
interacting dynamics, similar to that of the Chameleon mechanism
\cite{Khoury:2003rn}, or in a more general scenario to the symmetron cosmology
\cite{Hinterbichler:2011ca}. Another theoretical approach is given by
multi-scalar field theories, which have been proposed to unify the dark sector
\cite{Paliathanasis:2023moe,Luongo:2025ovo}.

In this context, within the formalism of symmetric teleparallel theory, we
introduce a coupling function to generate interacting dynamic in the dark
sector. In particular in the following, we examine the matter-gravity coupling
within the framework $f(Q)$-gravity can explain the late-time observational
data and the expansion history. In contrast with the previous analysis
presented in \cite{Paliathanasis:2025hjw,Abebe:2025wos}, in this investigation
we select a different connection for the definition of the gravitational
field, consequently, the definitions of the nonmetricity tensor and scalar. We
remark that in a FLRW geometry there are four different families of
connections \cite{Zhai:2023yny,Hohmann:2021ast,Dimakis:2022rkd}, leading to
four distinct gravitational theories in the context of $f\left(  Q\right)
$-gravity. Three of these families describe spatially flat FLRW geometries,
while the fourth family describes FLRW geometry with nonzero spatial
curvature. However, from the different gravitational theories related to the
three families of spatially flat geometry, only one gravitational model can be
recovered from that of the nonzero FLRW theory by eliminating the spatial
curvature \cite{Dimakis:2022rkd}. This specific connection is considered in
this study, since it is the only connection which allows us to have a
gravitational model with nonzero spatial curvature
\cite{Paliathanasis:2023raj}, or anisotropies \cite{Dimakis:2023uib} in the
very early stages of the universe.

Section \ref{sec2} provides a brief introduction to symmetric teleparallel
gravity and to $f\left(  Q\right)  $-gravity with matter-gravity coupling. In
Section \ref{sec3} we present the cosmological field equations for our model
which is that of a spatially flat FLRW geometry in power-law $f\left(
Q\right)  $-gravity with a pressureless fluid source coupled to gravity, where
we discuss the nonuniqueness in the selection of the connection for the
spatially flat universe. We make the choice for a noncoincidence connection.
Specifically, the connection that we select follows naturally from the unique
connection in the more general scenario of the presence of nonzero spatial
curvature in the background geometry.

The main findings of this analysis are presented in Section \ref{sec4}. We
present the analysis of the numerical results of the observational constraints
for the noncoincidence power-law $f\left(  Q\right)  $-gravity as dark energy
candidate by using differnt combindations of the late-time observational data.
Furthermore, we compare the statistical significant of the model in comparison
with the $\Lambda$CDM. Our results are summarized in Section \ref{sec5}.

\section{Symmetric teleparallel gravity}

\label{sec2}

We introduce the four-dimensional non-Riemannian manifold $M^{4}$ embedded
with the second-rank tensor $g_{\mu\nu}$, which plays the role of the metric.
The autoparallels in the manifold $M^{4}$ are defined by the connection
$\Gamma_{~\mu\nu}^{\lambda}~$. The selection of the connection rules the
geometric structure of the manifold.

In order to understand the geometric properties, we define the following
tensors%
\begin{align}
R_{\;\lambda\mu\nu}^{\kappa}  &  \equiv\Gamma_{\;\lambda\nu,\mu}^{\kappa
}-\Gamma_{\;\lambda\mu,\nu}^{\kappa}+\Gamma_{\;\lambda\nu}^{\sigma}%
\Gamma_{\;\mu\sigma}^{\kappa}-\Gamma_{\;\lambda\mu}^{\sigma}\Gamma
_{\;\mu\sigma}^{\kappa},\\
\mathrm{T}_{\mu\nu}^{\lambda}  &  \equiv\Gamma_{\;\mu\nu}^{\lambda}%
-\Gamma_{\;\nu\mu}^{\lambda},\\
Q_{\lambda\mu\nu}  &  \equiv g_{\mu\nu,\lambda}-\Gamma_{\;\lambda\mu}^{\sigma
}g_{\sigma\nu}-\Gamma_{\;\lambda\nu}^{\sigma}g_{\mu\sigma},
\end{align}
where $R_{\;\lambda\mu\nu}^{\kappa}$ is the Riemann tensor, $\mathrm{T}%
_{\mu\nu}^{\lambda}$ is the torsion tensor and $Q_{\lambda\mu\nu}$ correspond
to the nonmetricity tensor.

In General Relativity, the connection is considered to be the Levi-Civita,
from which leads to $\mathrm{T}_{\mu\nu}^{\lambda}=0$ and $Q_{\lambda\mu\nu
}=0$, where the gravitational field is derived by the Ricciscalar $R$. On the
other hand, in the framework of teleparallel gravity, the foundamental
geometric structure is given by the vierbein fields, which lead to
$R_{\;\lambda\mu\nu}^{\kappa}=0$, $Q_{\lambda\mu\nu}=0$, and gravity is
defined by the torsion scalar $\mathrm{T}$.

The introduction of a connection which is symmetric, i.e. $\mathrm{T}_{\mu\nu
}^{\lambda}=0$, and flat, i.e. $R_{\;\lambda\mu\nu}^{\kappa}=0$, leads to the
symmetric teleparallel theory, where the nonmetricity scalar $Q$ describes the
gravitational field.

Before we define the nonmetricity scalar we need to introduce the nonmetricity
conjugate tensor
\begin{equation}
P_{~\mu\nu}^{\lambda}=\frac{1}{4}\left(  -2L_{~~\mu\nu}^{\lambda}+Q^{\lambda
}g_{\mu\nu}-Q^{\prime\lambda}g_{\mu\nu}-\delta_{(\mu}^{\lambda}Q_{\nu
)}\right)  ,
\end{equation}
where $L_{~\mu\nu}^{\lambda}$ is the disformation parameter%
\begin{equation}
L_{~\mu\nu}^{\lambda}=\frac{1}{2}g^{\lambda\sigma}\left(  Q_{\mu\nu\sigma
}+Q_{\nu\mu\sigma}-Q_{\sigma\mu\nu}\right)  ,
\end{equation}
and $Q_{\lambda}=Q_{\lambda~~~\mu}^{~~~\mu},Q_{\lambda}^{\prime}%
=Q_{~~\lambda\mu}^{\mu}$ such that%
\begin{equation}
Q=Q_{\lambda\mu\nu}P^{\lambda\mu\nu}.
\end{equation}

In symmetric teleparallel gravity the Action Integral which replaces the
Einstein-Hilbert action is
\begin{equation}
S_{STGR}=\int d^{4}x\sqrt{-g}\left(  Q+\lambda_{\kappa}^{\;\lambda\mu\nu
}R_{\;\lambda\mu\nu}^{\kappa}+\tau_{\lambda}^{\;\mu\nu}\mathrm{T}_{\;\mu\nu
}^{\lambda}\right)  . \label{ac.01}%
\end{equation}
The Lagrange multipliers $\lambda_{\kappa}^{\;\lambda\mu\nu}$ and
$\tau_{\kappa}^{\;\lambda\mu\nu}$ have been introduced to enforce the flatness
and torsionless conditions for the connection.

There exist an algebraic relation for the scalar $Q$ and the Ricciscalar
$\mathring{R}$ is the curvature scalar of the metric tensor $g_{\mu\nu}$.
These two scalars are related by the formula
\begin{equation}
Q=\mathring{R}+\mathring{\nabla}_{\lambda}\left(  Q^{\lambda}-Q^{\prime
\lambda}\right)  ,
\end{equation}
where $\mathring{\nabla}_{\lambda}$ describes the covariant derivative in
terms of the Levi-Civita connection. Thus, by replacing the latter expression
in the action integral (\ref{ac.01}) we end with the Einstein-Hilbert action,
because $\mathring{\nabla}_{\lambda}\left(  Q^{\lambda}-Q^{\prime\lambda
}\right)  $ is a topological boundary term. Therefore, the Action Integral
(\ref{ac.01}) leads to the same equations of that of the Einstein-Hilbert action.

\subsection{$f\left(  Q\right)  $-gravity}

The introduction of nonlinear coefficients of the nonmetricity scalar in the
Action Integral (\ref{ac.01}) leads to the family of $f\left(  Q\right)
$-theories, similarly to the $f\left(  R\right)  $ generalization of General Relativity.

The gravitational Action Integral of $f\left(  Q\right)  $-theory is
\cite{BeltranJimenez:2019tme}
\begin{equation}
S_{f\left(  Q\right)  }=\int d^{4}x\sqrt{-g}\left(  f\left(  Q\right)
\right)  -\int d^{4}x\sqrt{-g}\mathcal{L}_{M}. \label{fq0}%
\end{equation}
where $\mathcal{L}_{M}$ describes contribution for the matter source.

Variation of (\ref{fq0}) with respect to the metric tensor leads to the
gravitational field equations \cite{BeltranJimenez:2019tme}%
\begin{equation}
\frac{2}{\sqrt{-g}}\nabla_{\lambda}\left(  \sqrt{-g}f_{,Q}P_{\mu\nu}^{\lambda
}\right)  -\frac{1}{2}f(Q)g_{\mu\nu}+f_{,Q}\left(  P_{\mu\rho\sigma}Q_{\nu
}^{\;\rho\sigma}-2Q_{\rho\sigma\mu}P_{\phantom{\rho\sigma}\nu}^{\rho\sigma
}\right)  =T_{\mu\nu}.
\end{equation}

The field equations read equivalently Einstein's tensor $G_{\mu\nu}%
=\mathring{R}_{\mu\nu}-\frac{\mathring{R}}{2}g_{\mu\nu}$, as follows
\cite{Zhai:2023yny}
\begin{equation}
f^{\prime}(Q)G_{\mu\nu}+\frac{1}{2}g_{\mu\nu}\left(  f_{,Q}Q-f(Q)\right)
+2f_{,QQ}\left(  \nabla_{\lambda}Q\right)  P_{\;\mu\nu}^{\lambda}=T_{\mu\nu}.
\label{feq1}%
\end{equation}
This expression allows to perform a direct comparison of the theory with the
linear limit of STEGR/GR. \ 

We introduce the effective energy momentum tensor $T_{\mu\nu}^{f\left(
Q\right)  }$ which attributes the geometric degrees of freedom related to the
modified theory. $T_{\mu\nu}^{f\left(  Q\right)  }$ is defined as
\begin{equation}
T_{\mu\nu}^{f\left(  Q\right)  }=-\left[  \frac{1}{2}g_{\mu\nu}\left(
f_{,Q}Q-f(Q)\right)  +2f_{,QQ}\left(  \nabla_{\lambda}Q\right)  P_{\;\mu\nu
}^{\lambda}\right]  .
\end{equation}
Consequently, the gravitational field equations (\ref{feq1}) are expressed in
the compact form
\begin{equation}
f_{,Q}G_{\mu\nu}=T_{\mu\nu}^{f\left(  Q\right)  }+T_{\mu\nu}. \label{feq2a}%
\end{equation}
Furthermore, variation of the Action Integral for the the connection leads to
the constraint
\begin{equation}
\nabla_{\mu}\nabla_{\nu}\left(  \sqrt{-g}f_{,Q}P_{\mu\nu}^{\lambda}\right)
=0. \label{feq2}%
\end{equation}
When the latter condition is trivially satisfied, we refer to the chosen
connection as the coincidence connection.

\subsection{Matter-Gravity Coupling}

The field equations (\ref{feq2a}) show that the theory provides a varying
gravitational constant which depends on the function $f\left(  Q\right)  $.
Nevertheless, in this work we introduce a coupling between the matter field
and the nonmetricity scalar in order to the resulting field equations to have
a constant gravitational constant.

In particular we modify the gravitational action integral (\ref{fq0}) as
follows \cite{Abebe:2025wos}
\begin{equation}
S_{f(Q)}=\int d^{4}x\sqrt{-g}\left[  f(Q)-\alpha f_{,Q}(Q)\mathcal{L}%
_{M}\right]  , \label{kd.04}%
\end{equation}
where $\alpha$ is the coupling parameter. Such interactions have been
investigated earlier within the context of other modified gravity theories.
\cite{Bertolami:2007gv,Bertolami:2008ab,Haghani:2021fpx}. In this approach,
the coupling function has been considered such that to eliminate the effects
of the varying gravitational constant. Now, the resulting gravitational field
equations are%
\begin{equation}
G_{\mu\nu}=\bar{T}_{\mu\nu}^{f\left(  Q\right)  }+T_{\mu\nu}, \label{kd.05}%
\end{equation}
where
\begin{equation}
\bar{T}_{\mu\nu}^{f\left(  Q\right)  }=\frac{1}{f_{,Q}}T_{\mu\nu}^{f\left(
Q\right)  }.
\end{equation}

As it has been found before in \cite{Abebe:2025wos}, this consideration can
overpass various problems within the framework of $f\left(  Q\right)
$-gravity. For instance, the conditions $f^{\prime}(Q)>0$ is not required
anymore for the validity of the gravitational model.

\section{FLRW Cosmology}

\label{sec3}

According to the Cosmological Principle, we consider a homogeneous and
isotropic geometry expressed by the spatially flat FLRW line element
\begin{equation}
ds^{2}=-N^{2}(t)\,dt^{2}+a^{2}(t)\left(  dx^{2}+dy^{2}+dz^{2}\right)  ,
\label{kd.02}%
\end{equation}
in which $N\left(  t\right)  $ represent the lapse function and $a(t)$ denotes
the scale factor which describe the radius of the three-dimensional hypersurface.

For the comoving observer $u^{\mu}=\frac{1}{N\left(  t\right)  }\delta
_{t}^{\mu},~$i.e. $u^{\mu}u_{\mu}=-1$, the expansion rate is defined as
$\theta=3H$, where now $H=\frac{1}{N}\frac{\dot{a}}{a}$ is the Hubble
function, and an overdot means differentiation with respect to the time variable.

For the line element (\ref{kd.02}), and within symmetric teleparallel theory,
the requirement that the connection to be flat, symmetric and inherits the
symmetries of the background space lead to three different connections, we
shall refer to them as $\Gamma^{A}$,~$\Gamma^{B}$ and $\Gamma^{C}$
\cite{Paliathanasis:2023pqp}.

The additional nonzero coefficients for connection $\Gamma^{A}$ are
\begin{equation}
\Gamma^{A}:\Gamma_{\;tt}^{t}=\gamma\left(  t\right)  .
\end{equation}
For this connection, condition (\ref{feq2}) holds trivially, thus, $\Gamma
^{A}$ is the coincidence connection. Therefore, function $\gamma\left(
t\right)  $ plays no role in the cosmological dynamics.

Connections $\Gamma^{B}$ and $\Gamma^{C}$ are the two noncoincidence
connections, in which the the corresponding nonzero components in the
coordinate system of the line element (\ref{kd.02}) are
\begin{equation}
\Gamma^{B}:\Gamma_{\;tt}^{t}=\frac{\dot{\psi}(t)}{\psi(t)}+\dot{\psi}%
(t),\quad\Gamma_{\;tx}^{x}=\Gamma_{\;ty}^{y}=\Gamma_{\;tz}^{z}=\dot{\psi
}\left(  t\right)  ,
\end{equation}%
\begin{equation}
\Gamma^{C}:\Gamma_{\;tt}^{t}=-\frac{\ddot{\Psi}}{\dot{\Psi}}~,\quad
\Gamma_{\;xx}^{t}=\Gamma_{yy}^{t}=\Gamma_{zz}^{t}=\frac{1}{\dot{\Psi}}.
\end{equation}
Condition (\ref{kd.02}) rules the dynamical motion for the scalars
$~\psi\left(  t\right)  $ and $\Psi\left(  t\right)  $.

In contrary to the dynamical variable $\gamma\left(  t\right)  $ of the
connection $\Gamma^{A}$,~$\Gamma^{B}$, for the remain two connections the
constraint (\ref{feq2}) is not trivially satisfied, and the scalars
$\psi\left(  t\right)  $ and $\Psi\left(  t\right)  $ have a nonzero
contribution in the evolution of the geometric field
\cite{Paliathanasis:2023pqp}.

For each connection we find the corresponding nonmetricity scalar as expressed
below are \cite{Paliathanasis:2023pqp}
\begin{align}
Q\left(  \Gamma^{A}\right)   &  =-6H^{2},\\
Q\left(  \Gamma^{B}\right)   &  =-6H^{2}+3\left(  3H\frac{\dot{\psi}}{N}%
+\frac{1}{N}\frac{d}{dt}\left(  \frac{\dot{\psi}}{N}\right)  ~~\right)  ,\\
Q\left(  \Gamma^{C}\right)   &  =-6H^{2}+\frac{3}{a^{2}}\left(  \frac{H}%
{\dot{\Psi}}+\frac{1}{N}\frac{d}{dt}\left(  \frac{1}{\dot{\Psi}N}\right)
\right)  .
\end{align}

The three scalars are connected through boundary terms and therefore describe
the same gravitational theory within STEGR, that is, $f\left(  Q\right)  $ be
a linear function. Nevertheless, this property is not valid for an arbitrary
function $f\left(  Q\right)  $. Each nonmetricity scalar provides a different
gravitational model and the choice of the connection is essential for the
physical properties of geometry.

We emphasize that, in the case of an FLRW geometry with nonvanishing spatial
curvature, within the symmetric teleparallel theory, there exists a unique
connection, namely $\Gamma^{C}$, defined in the noncoincidence gauge, which
satisfies the requirements of the theory.

Therefore, in the following we adopt the connection $\Gamma^{C}$ to describe
the gravitational dynamics. $\Gamma^{C}$ is the unique connection, among the
three, where the gravitational field equations for the flat universe can be
derived from the field equations of the FLRW geometry with nonzero curvature
within $f\left(  Q\right)  $-gravity, by eliminating coefficients related to
the spatial curvature. This property makes $\Gamma^{C}$ the only viable choice
for the consistent formulation of FLRW cosmology with or without spatial
curvature in symmetric teleparallel theory.

Similar properties of this connection have also been identified in homogeneous
and anisotropic Bianchi geometries, as well as in symmetric teleparallel
models describing astrophysical compact objects.
\cite{DAmbrosio:2021zpm,Dimakis:2024fan}.

For connection $\Gamma^{C}$ and for the matter source to be that of ideal gas
with zero pressure, which attributes the dust fluid components of the
universe, the Lagrangian function reads $\mathcal{L}_{M}=\rho_{m0}a^{-3}$.

Therefore, the modified Friedmann's equations are%
\begin{align}
3H^{2}-\frac{3}{2}a^{-2}\frac{\dot{\phi}}{\phi\dot{\Psi}}-\frac{V\left(
\phi\right)  }{\phi}-\rho_{m0}a^{-3}  &  =0,\label{f1}\\
\frac{2}{N}\dot{H}+3H^{2}+\frac{2}{N}H\frac{\dot{\phi}}{\phi}-\frac{1}%
{2}a^{-2}\frac{\dot{\phi}}{\phi\dot{\Psi}}+\frac{V(\phi)}{\phi}  &
=0,\label{f2}\\
\frac{1}{N}\left(  aN\frac{\dot{\phi}}{\dot{\Psi}^{2}}\right)  ^{\cdot}  &
=0,\label{f3}\\
-\frac{1}{a^{3}}\frac{1}{N}\left(  \frac{Na^{-2}}{\dot{\Psi}}\right)  ^{\cdot
}+2H^{2}-\frac{2}{3}V_{,\phi}  &  =-\frac{2}{3}\rho_{m0}a^{-3}, \label{f4}%
\end{align}
where the coupling coefficient $\alpha$ has been absorbed into $\rho_{m0}$,
the new scalar field $\phi$ attributes the dynamical degrees of freedom for
the $f\left(  Q\right)  $-gravity, that is,$~\phi=f_{,Q}$,~$V_{,\phi}=Q$, the
potential function is%
\begin{equation}
V\left(  \phi\right)  =Qf_{,Q}-f\left(  Q\right)  ,
\end{equation}
or
\begin{equation}
f(Q)=\phi(Q)V_{,\phi}(Q)-V(\phi(Q)).
\end{equation}
\ With this formalism we can use the minisuperspace description to write the
point-like Lagrangian that derives the gravitational field equations. The
scalar field description of $f\left(  Q\right)  $-gravity has been discussed
before, and it reveal the existence of minisuperspace description for the
gravitational model. In particular, the cosmological field equations
(\ref{f1})-(\ref{f4}) are given from the variation of the Action Integral
$S_{C}=\int\mathcal{L}_{C}\left(  N,a,\dot{a},\phi,\dot{\phi}\right)  dt$, in
which $\mathcal{L}_{C}$ is defined as \cite{Paliathanasis:2023pqp}
\begin{equation}
\mathcal{L}_{C}\left(  N,a,\dot{a},\phi,\dot{\phi}\right)  =-\frac{3}{N}\phi
a\dot{a}^{2}-\frac{3}{2}aN\frac{\dot{\phi}}{\dot{\Psi}}+Na^{3}V(\phi
)-N\rho_{m0}\phi.
\end{equation}

Regarding the $f\left(  Q\right)  $ model, we select the simplest nonlinear
functional forms which introduces the minimum number of new parameters in the
cosmological model.

We select the function%
\begin{equation}
f\left(  Q\right)  =f_{0}Q^{n},~n>1,
\end{equation}
where the limit $n=1$ corresponds to the STEGR/GR.

Thus, for this given $f\left(  Q\right)  $ theory, the corresponding potential
function is derived
\begin{equation}
V\left(  \phi\right)  =V_{0}\phi^{\frac{n}{n-1}},~V_{0}=V_{0}\left(
f_{0},n\right)  . \label{f5}%
\end{equation}

In what follows, we consider this power-law model as a framework for
explaining late-time observational data.

\section{Cosmological Observations}

\label{sec4} The observational constraints are presented in this Section. In
the following, we summarize the data sets considered in this work and the
methodology adopted for the statistical analysis.

\subsection{Late-time observations}

We consider the late-time observational data, with redshift $z<2.5$.
Specifically, we consider three different catalogues for the supernova (SNIa)
data, the Baryonic Acoustic Oscillations (BAO) from DESI DR2 collaboration and
the Observable Hubble Data (OHD) from the Cosmic Chronometers.

\begin{itemize}
\item SNIa: For our study we consider the PantheonPlus (PP)
\cite{Brout:2022vxf}, the Union3.0 (U3) \cite{rubin2023union} and the
DES-Dovekie (DESD) \cite{DES:2025sig} catalogues. These catalogues relate
observable values for the distance modulus $\mu^{obs}$ at the the redshift of
the event. The PP catalogue includes 1550 SNIa events within the redshifts
$10^{-3}<z<2.27$. Furthremore, the U3 catalogue is formed from 2087 events
within the same redshift as the PP data. U3 and PP catalogues share 1363 SNIa
events However, the photometric data analysis differs between PP and U3
leading to distinct catalogues. The reanalysis of five years of Type Ia
supernova data from the Dark Energy Survey (DES-SN5YR) led to the DD
catalogue, which includes 1820 SNIa events in the low-redshift regime
$z<1.13.$

\item BAO: We consider the latest BAO data from the Dark Energy Spectroscopic
Instrument (DESI DR2) \cite{DESI:2025zpo,DESI:2025zgx,DESI:2025fii} which
provides measurements of the transverse comoving angular distance ratio,
$\frac{D_{M}}{r_{drag}}=\frac{D_{L}}{\left(  1+z\right)  r_{drag}},~$the
volume-averaged distance ratio, $\frac{D_{V}}{r_{drag}}=\frac{\left(
zD_{H}D_{M}^{2}\right)  ^{1/3}}{r_{drag}}$ and the and the Hubble distance
ratio $\frac{D_{H}}{r_{d}}=\frac{c}{r_{drag}H(z)},~$at seven distinct
redshifts, where $~r_{drag}$ is the sound horizon at the baryon drag epoch.

\item OHD: We apply the thirty four data points from the Observable Hubble
Dataset (OHD) presented in \cite{moresco2020setting} and the three recent data
from the DESI DR1 observations \cite{Loubser:2025fzl}. These data follow from
the analysis of the cosmic chronometers. They provide model-independent direct
measurements of the Hubble parameter.
\end{itemize}

Table \ref{table0} shows the results corresponding to six distinct
configurations of the datasets described above.%

\begin{table}[tbp] \centering
\caption{Combinations of SNIa, BAO, and OHD datasets employed in this study.}%
\begin{tabular}
[c]{cccccc}\hline\hline
\textbf{Dataset} & \textbf{PP} & \textbf{U3} & \textbf{DD} & \textbf{BAO} &
\textbf{OHD}\\\hline
$\mathbf{D}_{1}$ & $\surd$ & $\times$ & $\times$ & $\surd$ & $\times$\\
$\mathbf{D}_{2}$ & $\times$ & $\surd$ & $\times$ & $\surd$ & $\times$\\
$\mathbf{D}_{3}$ & $\times$ & $\times$ & $\surd$ & $\surd$ & $\times$\\
$\mathbf{D}_{4}$ & $\surd$ & $\times$ & $\times$ & $\surd$ & $\surd$\\
$\mathbf{D}_{5}$ & $\times$ & $\surd$ & $\times$ & $\surd$ & $\surd$\\
$\mathbf{D}_{6}$ & $\times$ & $\times$ & $\surd$ & $\surd$ & $\surd
$\\\hline\hline
\end{tabular}
\label{table0}%
\end{table}%

\subsection{Methodology \& Priors}

We calculate the theoretical Hubble function from the field equations
(\ref{f1})-(\ref{f4}) by applying numerical techniques.

In particular, by introducing new dynamical variables through the Hubble
normalization approach \cite{Paliathanasis:2023nkb}, the field equations can
be expressed as a set of first-order differential equations.

We select the dynamical variables \cite{Paliathanasis:2023nkb}%
\begin{equation}
\Omega_{m}=\frac{\rho_{m0}}{3a^{3}H^{2}},~x_{\phi}=\frac{\dot{\phi}}{H\phi
},~x_{\Psi}=a^{2}H\dot{\Psi},~y=\frac{V(\phi)}{3H^{2}\phi}~,~\lambda
=\frac{\phi V_{,\phi}}{V}, \label{ss1}%
\end{equation}
and the independent variable $\tau=-\ln\left(  1+z\right)  $.

Therefore the dynamical system (\ref{f1})-(\ref{f4}) is expressed as follows%
\begin{align}
-\left(  1+z\right)  \frac{dx_{\phi}}{dz}  &  =\frac{1}{4}x_{\phi}\left(
10-x_{\phi}x_{\Psi}+6y-\frac{16\left(  1-\lambda y+\Omega_{m}\right)
}{x_{\Psi}}\right)  ,\label{ss.01}\\
-\left(  1+z\right)  \frac{dx_{\Psi}}{dz}  &  =\frac{1}{x_{\Psi}^{2}}\left(
x_{\Psi}\left(  6+x_{\phi}\left(  x_{\Psi}-4\right)  -6y\right)  +8\lambda
y-8\left(  1+\Omega_{m}\right)  \right)  ,\label{ss.02}\\
-\left(  1+z\right)  \frac{dy}{dz}  &  =y\left(  x_{\phi}\left(
1+\lambda-\frac{x_{\psi}}{2}\right)  +3\left(  1+y\right)  \right)
,\label{ss.03}\\
-\left(  1+z\right)  \frac{d\Omega_{m}}{dz}  &  =-\Omega_{m}x_{\phi}\left(
\left(  \frac{x_{\Psi}}{2}-2\right)  -3y\right)  ,\label{ss.04}\\
-\left(  1+z\right)  \frac{d\lambda}{dz}  &  =x_{\phi}g(\lambda),
\label{ss.05}%
\end{align}
with constraint%
\begin{equation}
\Omega_{m}=1-\frac{1}{2}x_{\phi}x_{\Psi}+y. \label{ss.06}%
\end{equation}
in which
\begin{equation}
g\left(  \lambda\left(  \phi\right)  \right)  =\lambda\frac{V_{,\phi\phi}%
}{V_{,\phi}}-\lambda^{2}+\lambda.
\end{equation}

For the gravitational model (\ref{f5}) we derive $\lambda=n$ and $g\left(
\lambda\right)  =0$. Therefore, by applying conditions (\ref{ss.06}) we
finally obtain the reduced three-dimensional system on the phase-space
$\left\{  x_{\phi},x_{\psi},y\right\}  $. Moreover, for the cosmological
solution, the deceleration parameter is expressed in terms of the new
variables from the algebraic expression
\begin{equation}
q\left(  z\right)  =\frac{1}{2}\left(  1+\frac{1}{2}x_{\phi}\left(  4-x_{\Psi
}\right)  +3y\right)  .
\end{equation}
Therefore, the Hubble function is given as
\begin{equation}
\frac{H\left(  z\right)  }{H_{0}}=\exp\left(  \int_{0}^{z}\frac{1+q\left(
z\right)  }{1+z}dz\right)  .
\end{equation}
%

\begin{table}[tbp] \centering
\caption{Priors of the parameters used for the MCMC sampler.}%
\begin{tabular}
[c]{cc}\hline\hline
\textbf{Parameters} & \textbf{Priors}\\\hline
$\mathbf{H}_{0}$ & $\left[  60,80\right]  $\\
$\mathbf{\Omega}_{m0}$ & $\left(  0,1\right)  $\\
$\mathbf{r}_{drag}$ & $\left[  130,160\right]  $\\
$\mathbf{x}_{\phi}^{0}$ & $\left[  -0.1,0.1\right]  $\\
$\mathbf{x}_{\Psi}^{0}$ & $\left(  0,35\right)  $\\
$\mathbf{\lambda}^{-1}$ & $\left(  0,1\right)  $\\\hline\hline
\end{tabular}
\label{table01}%
\end{table}%

We employ the Runge-Kutta numerical integration to solve the differential
equations (\ref{ss.01}), (\ref{ss.02}) and (\ref{ss.03}) with initial
conditions $\left\{  x_{\phi}^{0},x_{\Psi}^{0},y_{0}\right\}  $, where
$y_{0}=\frac{1}{2}x_{\phi}^{0}x_{\Psi}^{0}+\Omega_{m0}-1$. Parameter
$\Omega_{m0}$ denotes the energy density for the dust fluid source at the
present. We assume that $\Omega_{m0}$ attributes the dark matter and the
baryons. Parameter $\lambda$ is the only free parameter for the dynamical
system. If we select another functional form for the $f\left(  Q\right)
$-theory, then $g\left(  \lambda\right)  \neq0$, and $\lambda$ will be a
dynamical variable. Thus we would have as new parameters, the initial
condition $\lambda_{0}$, and any new parameter introduced by the $f\left(
Q\right)  $ function within the $g\left(  \lambda\right)  $.

The cosmological solution obtained from the numerical simulations is used to
compute observables, which we use to explain the datasets presented above.
Parameter estimation is performed via the Bayesian inference
COBAYA\footnote{https://cobaya.readthedocs.io/} \cite{cob1,cob2} employing the
MCMC sampler~\cite{mcmc1,mcmc2}. For the analysis of the MCMC chains we
utilize the GetDist library\footnote{https://getdist.readthedocs.io/}%
~\cite{getd}.

The free parameters for our model are the initial conditions $\left\{
x_{\phi}^{0},x_{\Psi}^{0},\Omega_{m0}\right\}  $, \ parameter $\lambda$, the
value of the Hubble parameter today, i.e. $H_{0}$ and $r_{drag}$ which is also
treated as a free parameter to be constrained. The priors used in this
investigation are presented in Table \ref{table01}.

\begin{figure}[t]
\centering\includegraphics[width=0.8\textwidth]{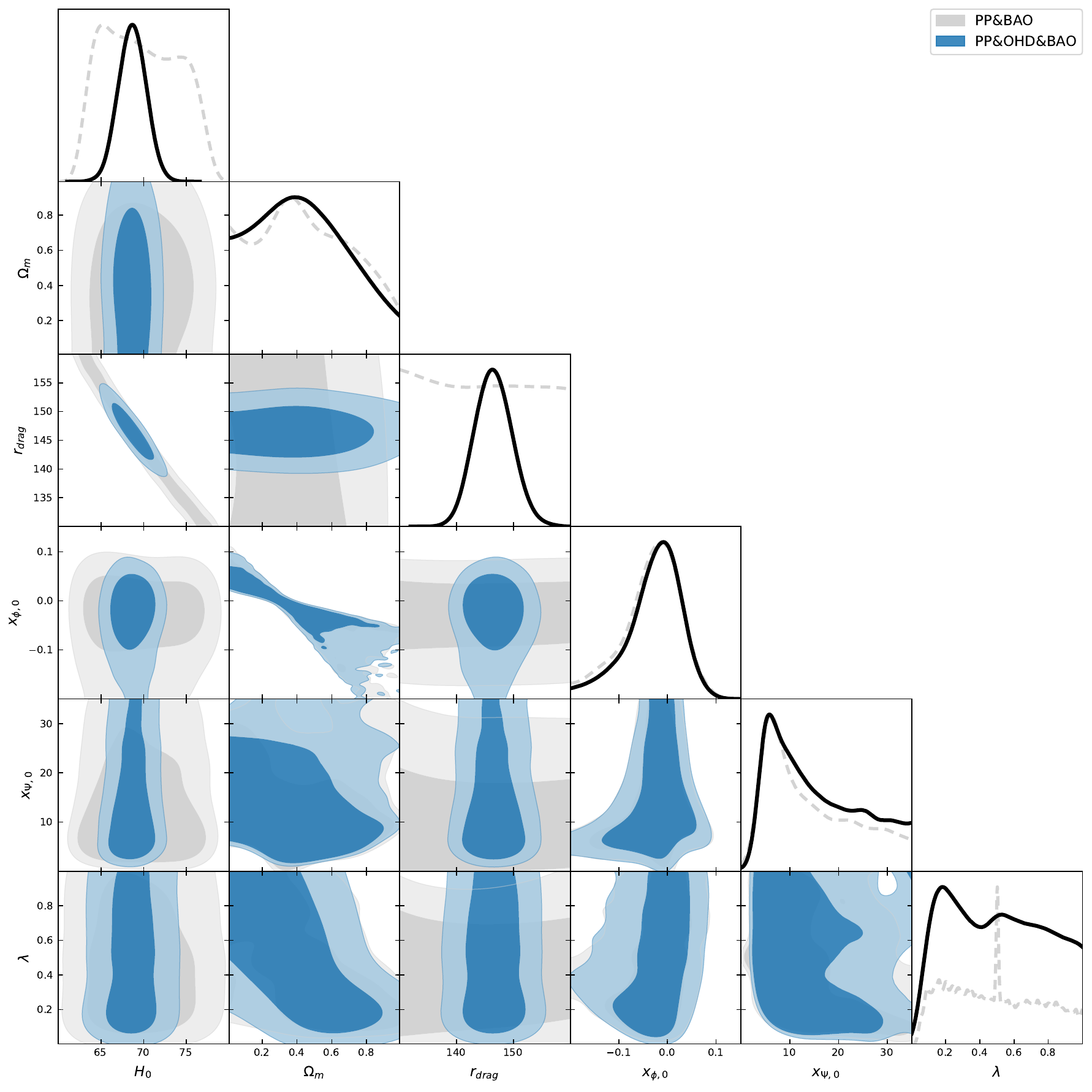}\caption{Confidence
regions of the numerical results of the non-coincidence $f\left(  Q\right)
$-gravity as derived from the datasets $D_{1}:PP\&BAO$ and $D_{4}%
:PP\&OHD\&BAO$.}%
\label{fig1}%
\end{figure}

\begin{figure}[t]
\centering\includegraphics[width=0.8\textwidth]{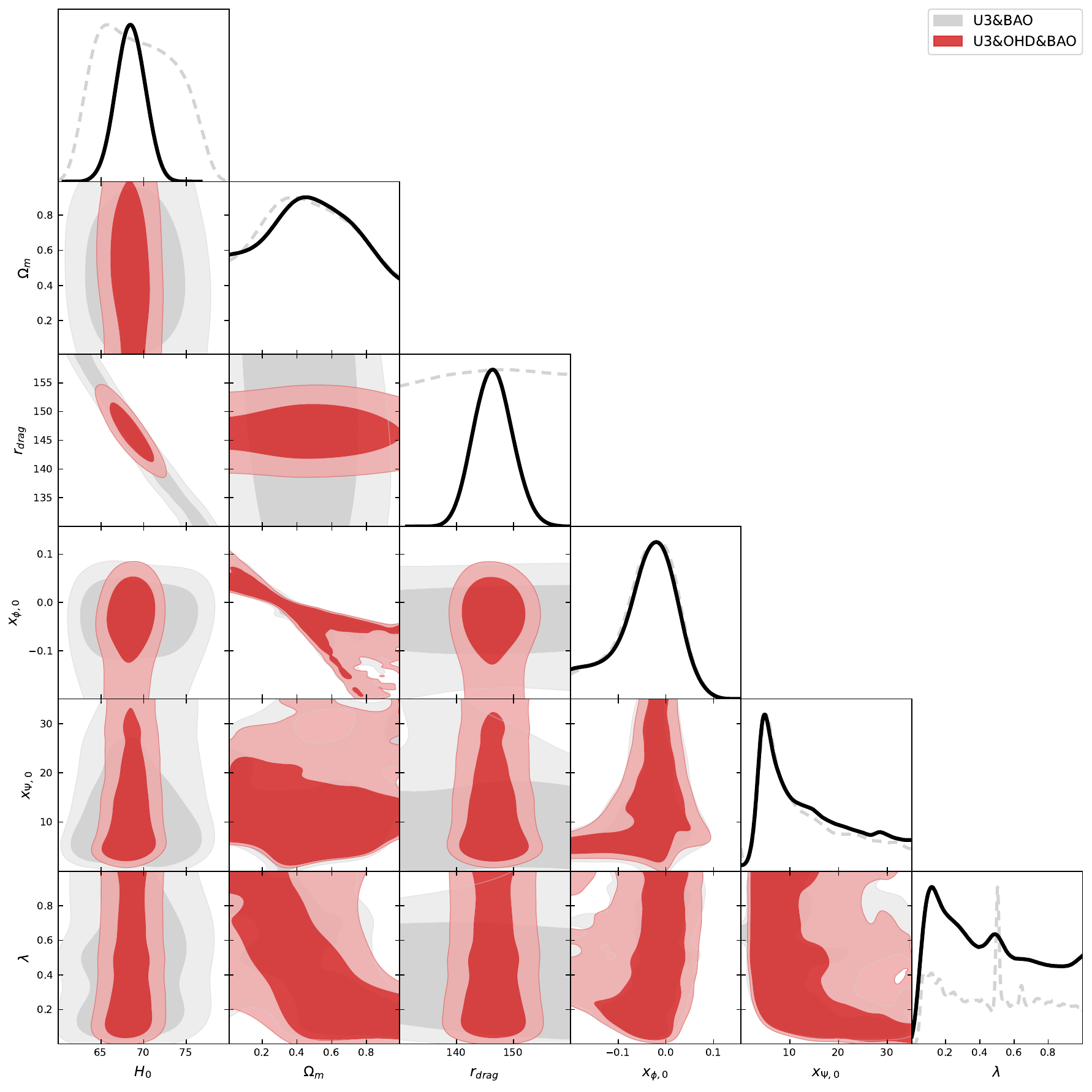}\caption{Confidence
regions of the numerical results of the non-coincidence $f\left(  Q\right)
$-gravity as derived from the datasets $D_{2}:U3\&BAO$ and $D_{5}%
:U3\&OHD\&BAO$.}%
\label{fig2a}%
\end{figure}

\begin{figure}[t]
\centering\includegraphics[width=0.8\textwidth]{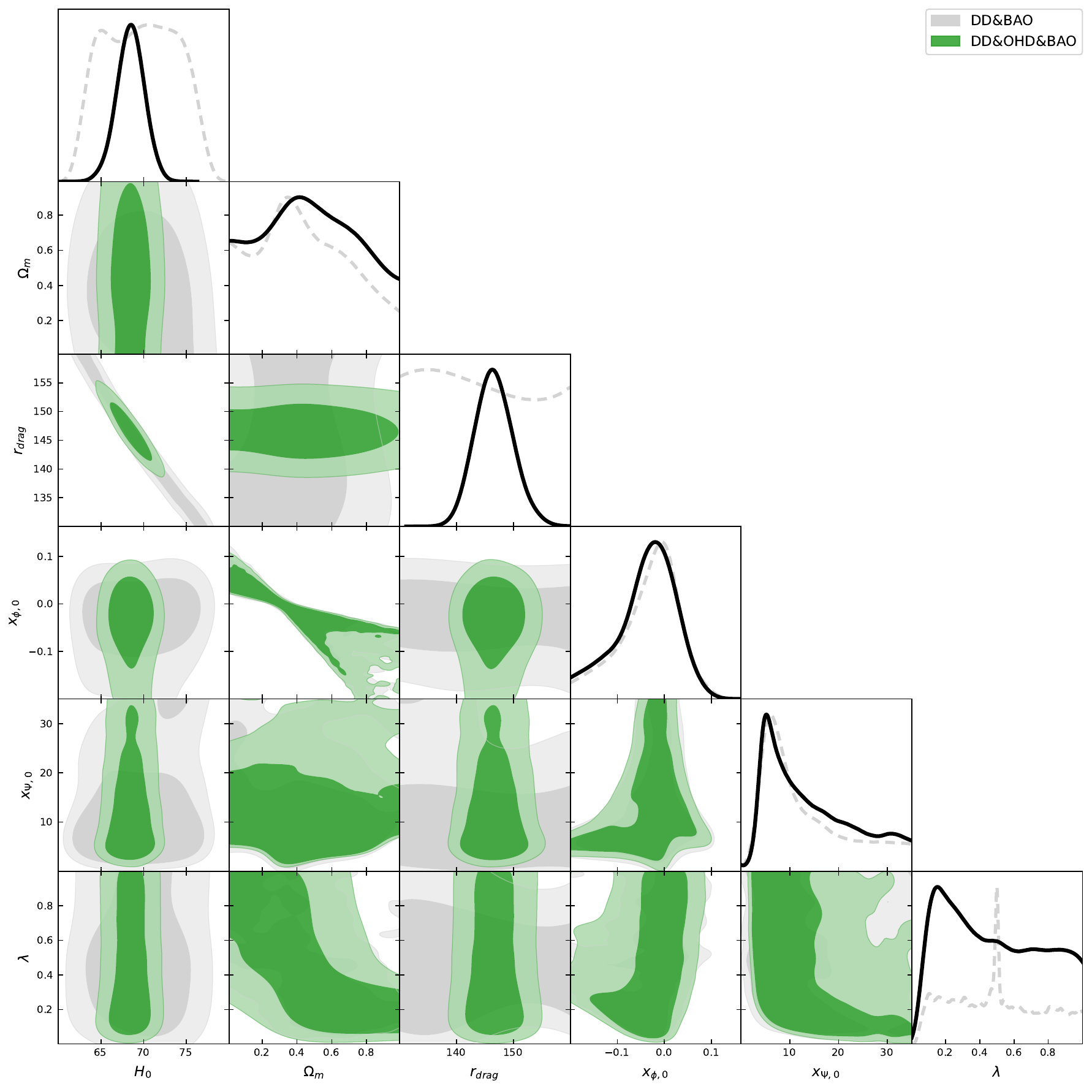}\caption{Confidence
regions of the numerical results of the non-coincidence $f\left(  Q\right)
$-gravity as derived from the datasets $D_{3}:DD\&BAO$ and $D_{6}%
:DD\&OHD\&BAO$.}%
\label{fig3a}%
\end{figure}

\subsection{Results}

With the GetDist library we examine the MCMC chains and we obtain the mean
values and the marginalized posterior credible intervals (CIs) at the 68\% and
95\% level as computed from the MCMC chains using the GetDist library for the
parameters $\left\{  H_{0},\Omega_{m0},x_{\phi}^{0},x_{\Psi}^{0},\lambda
^{-1},r_{drag}\right\}  $ as summarized in Tables \ref{table3a} and
\ref{table3b}. \ The difference between the mean and median values, together
with the asymmetry of the credible intervals reveal the non-Gaussian character
of the posteriors.

\begin{figure}[t]
\centering\includegraphics[width=0.8\textwidth]{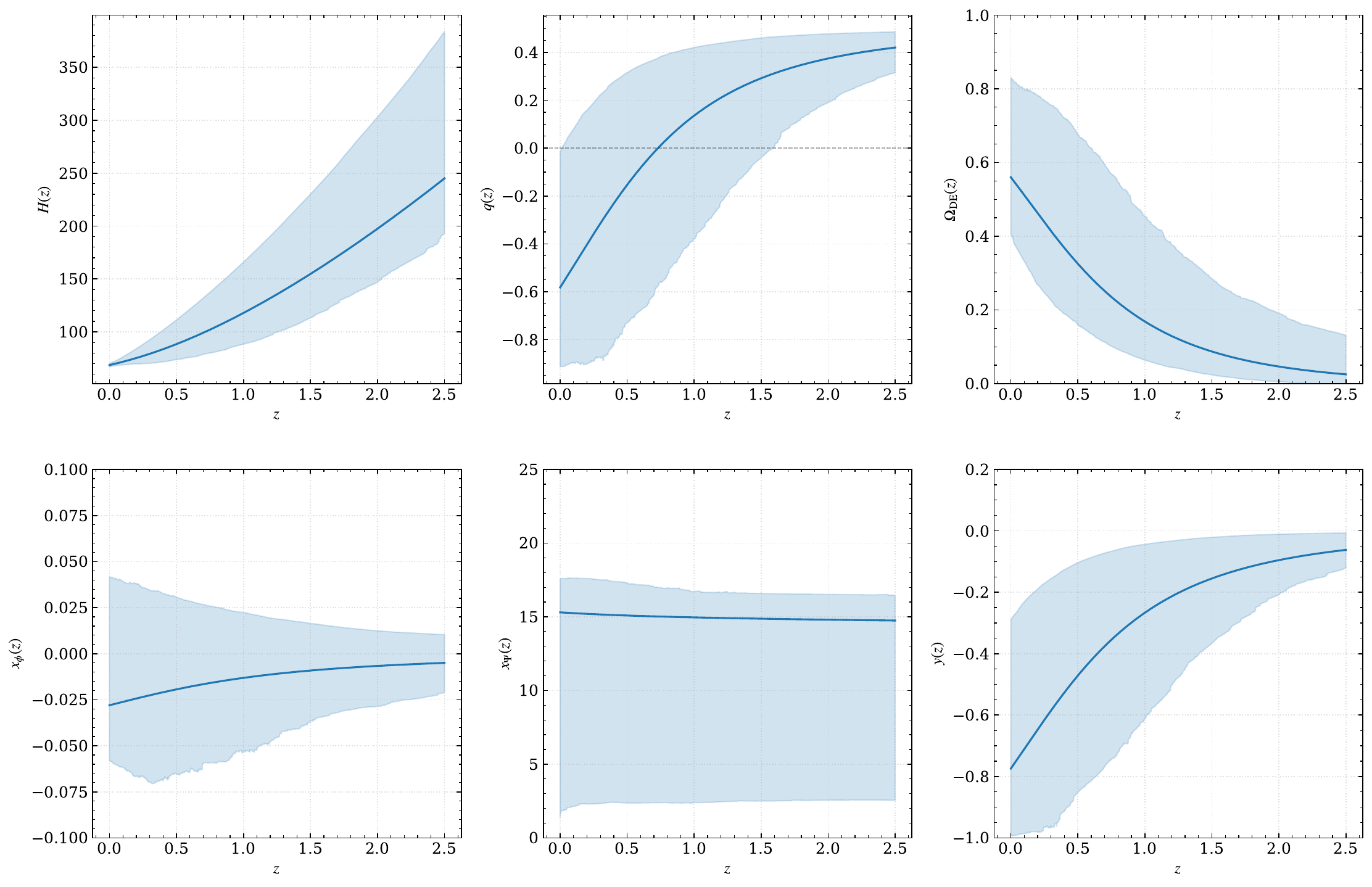}\caption{Parameter
space showing the dynamical evolution of the Hubble function $H\left(
z\right)  $, \ the deceleration parameter $q\left(  z\right)  $, effective
dark energy density $\Omega_{DE}\left(  z\right)  $, the kinetic term of the
scalar field $x_{\phi}\left(  z\right)  $, the connection component $x_{\Psi
}\left(  z\right)  $ and variable $y\left(  z\right)  $, for the parameter
values with the 68\% credible intervals derived from the observational
constraints of dataset $D_{4}.$}%
\label{fig4a}%
\end{figure}

\begin{figure}[t]
\centering\includegraphics[width=0.8\textwidth]{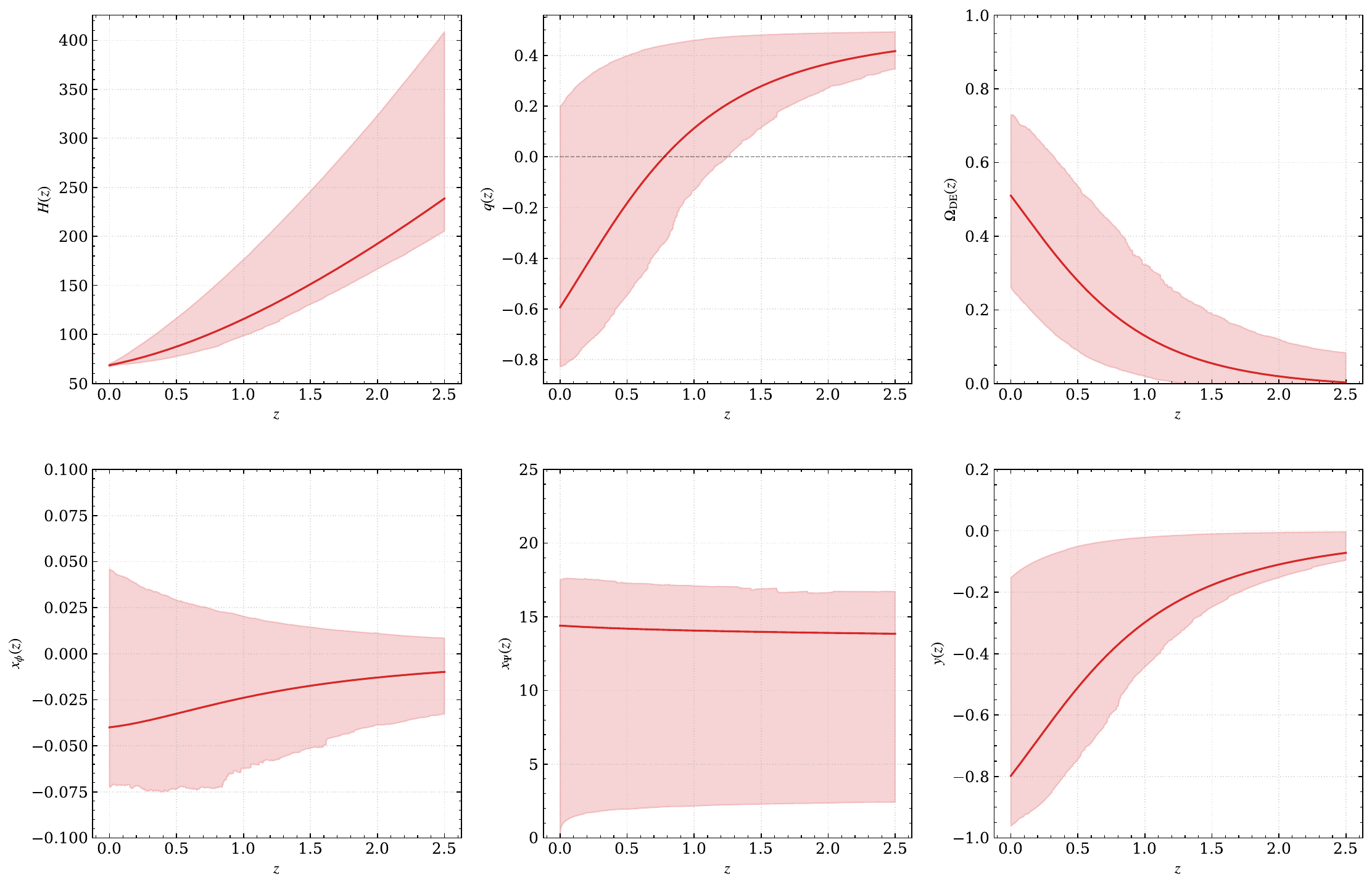}\caption{Parameter
space showing the dynamical evolution of the Hubble function $H\left(
z\right)  $, \ the deceleration parameter $q\left(  z\right)  $, effective
dark energy density $\Omega_{DE}\left(  z\right)  $, the kinetic term of the
scalar field $x_{\phi}\left(  z\right)  $, the connection component $x_{\Psi
}\left(  z\right)  $ and variable $y\left(  z\right)  $, for the parameter
values within the 68\% credible intervals derived derived from the
observational constraints of dataset $D_{5}.$}%
\label{fig5}%
\end{figure}

\begin{figure}[t]
\centering\includegraphics[width=0.8\textwidth]{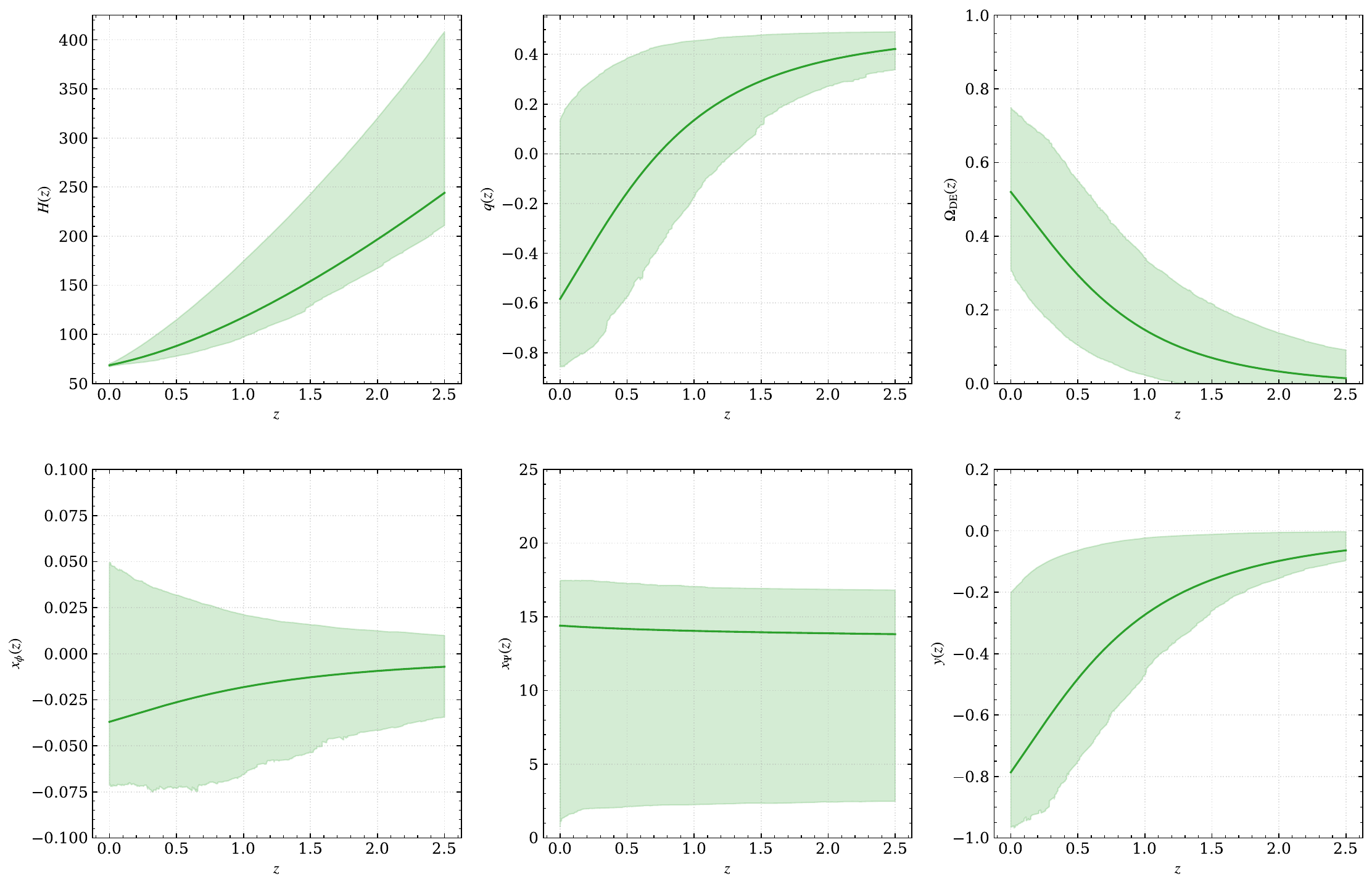}\caption{Parameter
space showing the dynamical evolution of the Hubble function $H\left(
z\right)  $, \ the deceleration parameter $q\left(  z\right)  $, effective
dark energy density $\Omega_{DE}\left(  z\right)  $, the kinetic term of the
scalar field $x_{\phi}\left(  z\right)  $, the connection component $x_{\Psi
}\left(  z\right)  $ and variable $y\left(  z\right)  $, for the parameter
values with the 68\% credible intervals derived derived from the observational
constraints of dataset $D_{6}.$}%
\label{fig6a}%
\end{figure}

In Figs. \ref{fig1}, \ref{fig2a} and \ref{fig3a}, we present the confidence
region of the cosmological parameters, while in Figs. \ref{fig4a}, \ref{fig5}
and \ref{fig6a} show the evolution of the cosmological quantities $H\left(
z\right)  $, $q\left(  z\right)  $, $\Omega_{DE}\left(  z\right)  $, as well
as the dynamical variables $x_{\phi}\left(  z\right)  ,~x_{\Psi}\left(
z\right)  $ and $y\left(  z\right)  $ for the parameter values and 68\%
credible intervals derived from the observational constraints of datasets
$D_{4}$,~$D_{5}$ and $D_{6}$ respectively.

The numerical solutions indicate that the variable related to the connection
$x_{\Psi}$ remains comparatively large and constant, while the scalar-field
kinetic contribution component $x_{\phi}$ is very small. This indicate that
the late-time acceleration is mainly driven by the potential term, that is,
variable $y$.%

\begin{table}[tbp] \centering
\caption{Observational constraints for the free parameters of noncoincidence $f(Q)$-gravity for the 68\% and 95\% credible intervals for the datasets with SNIa+BAO.}
\begin{tabular}
[c]{cccc}\hline\hline
\textbf{Dataset} & $\mathbf{D}_{1}$ & $\mathbf{D}_{2}$ & $\mathbf{D}_{3}%
$\\\hline
\multicolumn{4}{c}{\textbf{SNIa+BAO}}\\\hline
$\mathbf{H}_{0}$ & $69.7_{-5.9\left(  -6.9\right)  }^{+5.2\left(  +7.4\right)
}$ & $69.3_{-5.4\left(  -7.0\right)  }^{+3.7\left(  +7.5\right)  }$ &
$69.7_{-4.2\left(  -7.0\right)  }^{+4.2\left(  +7.0\right)  }$\\
$\mathbf{\Omega}_{m0}$ & $<0.596\left(  -\right)  $ & $0.48_{-0.32}%
^{+0.27}\left(  -\right)  $ & $<0.571\left(  -\right)  $\\
$\mathbf{r}_{drag}$ & $-$ & $-$ & $-$\\
$\mathbf{x}_{\phi}^{0}$ & $-0.032_{-0.042\left(  -0.14\right)  }%
^{+0.069\left(  +0.10\right)  }$ & $-0.039_{-0.043\left(  -0.14\right)
}^{+0.073\left(  +0.10\right)  }$ & $-0.032_{-0.045\left(  -0.14\right)
}^{+0.071\left(  +0.11\right)  }$\\
$\mathbf{x}_{\Psi}^{0}$ & $14.4_{-12\left(  -12\right)  }^{+7.3\left(
+18\right)  }$ & $13.6_{-11\left(  -11\right)  }^{+4.5\left(  +19\right)  }$ &
$13.3_{-11\left(  -11\right)  }^{+3.7\left(  +19\right)  }$\\
$\mathbf{\lambda}^{-1}$ & $0.48_{-0.26}^{+0.26}\left(  >0.10~\text{at}%
~95\%\right)  $ & $0.48_{-0.43\left(  -0.43\right)  }^{+0.36\left(
+0.51\right)  }$ & $0.49_{-0.26}^{+0.26}\left(  >0.10~\text{at}~95\%\right)
$\\\hline
$\mathbf{\chi}_{\min}^{2}-\mathbf{\chi}_{\min\Lambda}^{2}$ & \thinspace$-3.7$
& $\,-6.0$ & $-4.9$\\\hline\hline
\end{tabular}
\label{table3a}%
\end{table}%
%

\begin{table}[tbp] \centering
\caption{Observational constraints for the free parameters of noncoincidence $f(Q)$-gravity for the 68\% and 95\% credible intervals for the datasets with SNIa+OHD+BAO.}
\begin{tabular}
[c]{cccc}\hline\hline
\textbf{Dataset} & $\mathbf{D}_{4}$ & $\mathbf{D}_{5}$ & $\mathbf{D}_{6}%
$\\\hline
\multicolumn{4}{c}{\textbf{SNIa+OHD+BAO}}\\\hline
$\mathbf{H}_{0}$ & $68.7_{-1.6\left(  -3.2\right)  }^{+1.6\left(  +3.2\right)
}$ & $68.5_{-1.7\left(  -3.3\right)  }^{+1.7\left(  +3.3\right)  }$ &
$68.5_{-1.6\left(  -3.3\right)  }^{+1.6\left(  +3.2\right)  }$\\
$\mathbf{\Omega}_{m0}$ & $0.44_{-0.35}^{+0.19}\left(  -\right)  $ &
$0.49_{-0.26}^{+0.26}\left(  -\right)  $ & $0.48_{-0.26}^{+0.26}\left(
-\right)  $\\
$\mathbf{r}_{drag}$ & $146.5_{-3.3\left(  -6.2\right)  }^{+3.3\left(
+6.6\right)  }$ & $146.4_{-3.4\left(  -6.3\right)  }^{+3.4\left(  +6.7\right)
}$ & $146.6_{-3.6\left(  -6.3\right)  }^{+3.2\left(  +7.1\right)  }$\\
$\mathbf{x}_{\phi}^{0}$ & $-0.028_{-0.038\left(  -0.13\right)  }%
^{+0.064\left(  +0.10\right)  }$ & $-0.040_{-0.042\left(  -0.14\right)
}^{+0.074\left(  -0.10\right)  }$ & $-0.037_{-0.044\left(  -0.14\right)
}^{+0.074\left(  +0.11\right)  }$\\
$\mathbf{x}_{\Psi}^{0}$ & $15.3_{-12}^{+5.4}\left(  >3.97~\text{at}%
~95\%\right)  $ & $14.4_{-12\left(  -12\right)  }^{+4.9\left(  +18\right)  }$
& $14.4_{-11\left(  -12\right)  }^{+4.7\left(  +18\right)  }$\\
$\mathbf{\lambda}^{-1}$ & $0.50_{-0.40\left(  -0.41\right)  }^{+0.25\left(
+0.48\right)  }$ & $>0.27\left(  >0.07~\text{at}~95\%\right)  $ &
$0.48_{-0.28\left(  -0.41\right)  }^{+0.28\left(  +0.49\right)  }$\\\hline
$\mathbf{\chi}_{\min}^{2}-\mathbf{\chi}_{\min\Lambda}^{2}$ & $-3.4$ & $-5.8$ &
$-4.4$\\\hline\hline
\end{tabular}
\label{table3b}%
\end{table}%

The values of the free paramters obtained from the numerical simulations are
discussed below.

\begin{itemize}
\item $\mathbf{D}_{1}:$ The combination of the PP SNIa catalogue with the BAO
provide the best-fit parameters with the 68\% credible intervals
$H_{0}=69.7_{-5.9}^{+5.2}$,~$\Omega_{m0}<0.596$,~$x_{\phi}^{0}=-0.032_{-0.042}%
^{+0.069}$,~$x_{\psi}^{0}=14.4_{-12}^{+7.3}$ and $\lambda^{-1}=0.48_{-0.26}%
^{+0.26}$. Comparing the $\chi_{\min}^{2}$ with that obtained for the
$\Lambda$CDM, it follows $\mathbf{\chi}_{\min}^{2}-\mathbf{\chi}_{\min\Lambda
}^{2}=-3.7$.

\item $\mathbf{D}_{2}:$ Nevertheless, the consideration of the U3 catalogue
provides $H_{0}=69.7_{-5.9}^{+5.2}$,~$\Omega_{m0}=0.48_{-0.32}^{+0.27}%
$,~$x_{\phi}^{0}=-0.039_{-0.043}^{+0.073}$,~$x_{\psi}^{0}=13.6_{-11}^{+4.5}$
and $\lambda^{-1}=0.48_{-0.43}^{+0.36}$. \ We observe that now there is lower
limit for $\Omega_{m0}$. Furthermore, the comparison with the $\Lambda
$CDM~provides $\mathbf{\chi}_{\min}^{2}-\mathbf{\chi}_{\min\Lambda}^{2}=-6$.

\item $\mathbf{D}_{3}:$ The analysis of the MCMC chains from the combination
of DD SNIa events and the BAO data, gives $H_{0}=69.7_{-4.2}^{+4.2}$%
,~$\Omega_{m0}<0.571$,~$x_{\phi}^{0}=-0.032_{-0.045}^{+0.071}$,~$x_{\psi}%
^{0}=13.3_{-11}^{+3.7}$ and $\lambda^{-1}=0.49_{-0.26}^{+0.26}$, where again
the analysis provide a smaller $\chi_{\min}^{2}$ value from the $\Lambda$CDM,
that is, $\mathbf{\chi}_{\min}^{2}-\mathbf{\chi}_{\min\Lambda}^{2}=-4.9$.

\item $\mathbf{D}_{4}:$ The introduction of the OHD measurements with the PP
and BAO data leads to better constraints and smaller uncertainties. We find
$H_{0}=68.7_{-1.6}^{+1.6}$,~$\Omega_{m0}=0.44_{-0.35}^{+0.19}$,~$x_{\phi}%
^{0}=-0.028_{-0.038}^{+0.064}$,~$x_{\psi}^{0}=15.3_{-12}^{+5.4}$ and
$\lambda^{-1}=0.50_{-0.40}^{+0.25}$ and $\mathbf{\chi}_{\min}^{2}%
-\mathbf{\chi}_{\min\Lambda}^{2}=-3.4$.

\item $\mathbf{D}_{5}:$ Furthermore, the combination of data
U3\&OHD\&BAO\ provides, $H_{0}=68.5_{-1.7}^{+1.7}$,~$\Omega_{m0}%
=0.49_{-0.26}^{+0.26}$,~$x_{\phi}^{0}=-0.040_{-0.042}^{+0.074}$,~$x_{\psi}%
^{0}=14.4_{-12}^{+4.9}$ and $\lambda^{-1}>0.27$ and $\mathbf{\chi}_{\min}%
^{2}-\mathbf{\chi}_{\min\Lambda}^{2}=-5.8$.

\item $\mathbf{D}_{6}:$ Finally the study of the MCMC chains for the data set
DD\&OHD\&BAO leads to the parametric space $H_{0}=68.5_{-1.6}^{+1.6}$%
,~$\Omega_{m0}=0.48_{-0.26}^{+0.26}$,~$x_{\phi}^{0}=-0.037_{-0.044}^{+0.074}%
$,~$x_{\psi}^{0}=14.4_{-11}^{+4.7}$ and $\lambda^{-1}=0.48_{-0.28}^{+0.28}$
and $\mathbf{\chi}_{\min}^{2}-\mathbf{\chi}_{\min\Lambda}^{2}=-4.4$.
\end{itemize}

Overall the analysis suggests that for the different datasets there exist
consistent constraints on the cosmological parameters. $\Omega_{m0}$ is weakly
bounded with relative large uncertainties. The initial condition $x_{\phi}%
^{0}$ is close to zero, and $x_{\Psi}^{0}$ is relative large with large
uncertainty, however the combination $x_{\phi}^{0}x_{\Psi}^{0}$ is relative small.

As far as the power of gravity is concerned parameter $\lambda$ is constraint
to be around $\lambda\simeq2$, providing that the power-law $f\left(
Q\right)  $ model is the quadratic, i.e. $f\left(  Q\right)  =f_{0}Q^{2}$.
This is an interesting result, because the $Q^{2}$ has been derived before
\cite{Dimakis:2022wkj} in the analysis for the existence of scaling solutions
for the FLRW universe with or without spatial curvature.

\begin{figure}[t]
\centering\includegraphics[width=0.4\textwidth]{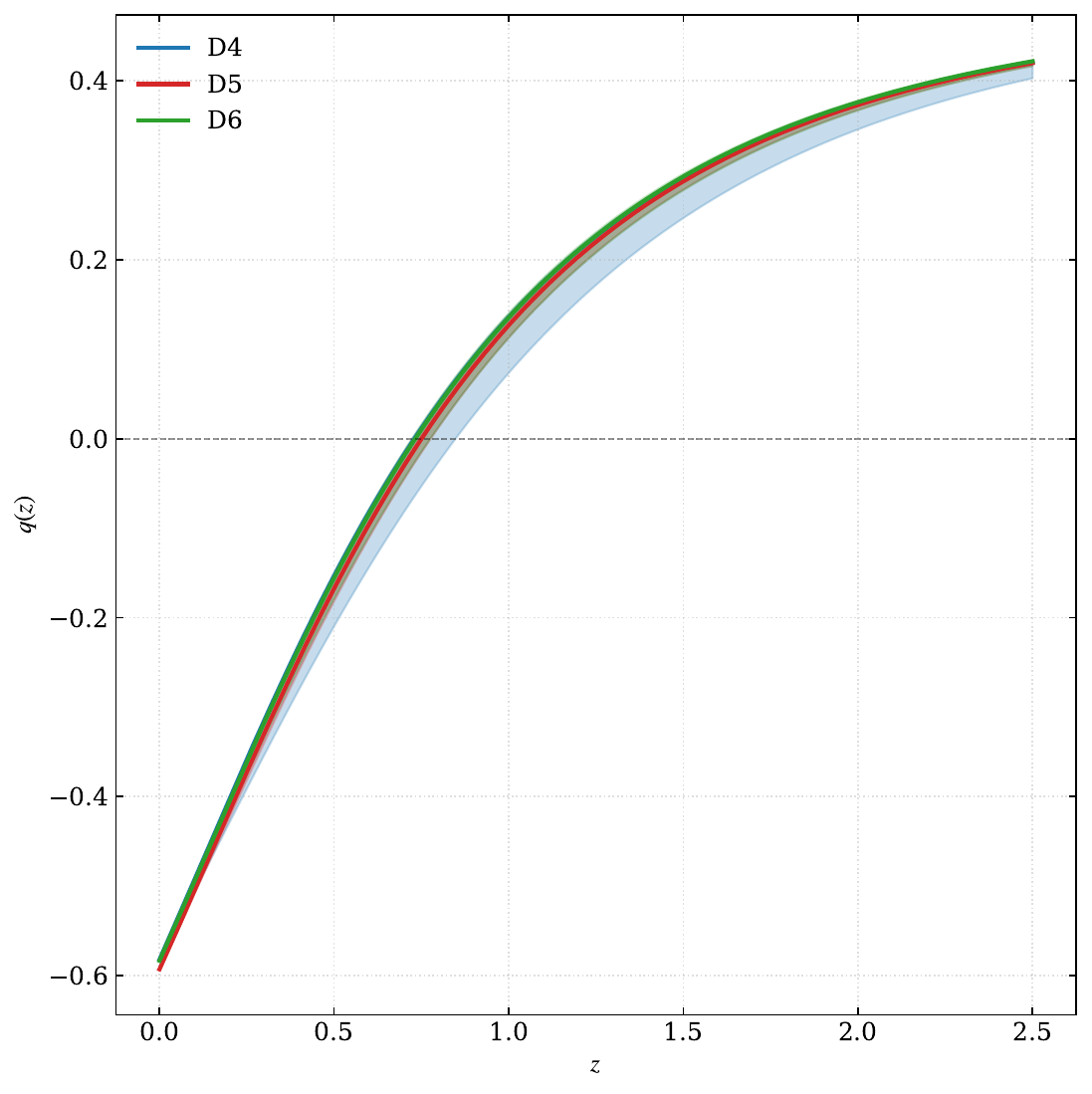}\caption{The
deceleration parameter $q\left(  z\right)  $, for the parameters as derived by
the datasets $D_{4}$,~$D_{5}$ and $D_{6}$ and varying parameter $\lambda$
within the 68\% credible intervals derived.}%
\label{fig7}%
\end{figure}

To explore the impact of the free parameters on the dynamics of the physical
quantities, we study the cosmological dynamics by fixing all parameters to
their best-fit values and varying only one parameter at a time, while keeping
the remaining parameters constant. In Fig. \ref{fig7} we present the evolution
of the deceleration parameter $q\left(  z\right)  $, for the best-fit
parameters of datasets $D_{4}$,~$D_{5}$ and $D_{6}$ by varying parameter
$\lambda$. We observe that $\lambda$ has a small impact on the behavior of the
deceleration parameter which is in agreement with the results obtained from
the analysis of the MCMC chains. Indeed, parameter $\lambda$ is weakly
constrains by the data. Recall that the lower constraint $\lambda>=1$, is
theoretically motivated, in order the theory to be well defined.

\subsection{Information Criterion}

The $\Lambda$CDM model and the power-law $f\left(  Q\right)  $-gravity~$\kappa
$, have different degrees dimension for the parametric space, $\Lambda$CDM has
three free parameters, while the $f\left(  Q\right)  $-gravity has six. Hence
the comparison of the $\chi_{\min}^{2}$ is not sufficient to conclude about
the statistical preferred model by the datasets.\ 

We employ the Akaike Information Criterion (AIC) \cite{AIC}, which allows us
to evaluate which model is supported by the late-time cosmological datasets,
by penalty models with large parametric space. We introduce the AIC parameter
where for large datasets it is expressed as%
\begin{equation}
{AIC}\simeq\chi_{\min}^{2}+2\kappa.
\end{equation}
which is used to compare the two model by applying Akaike's scale.

Indeed, the Akaike information criterion indicates which model provides a
better fit to the data by comparing the differences in their AIC values, i.e.
$\Delta AIC=AIC_{A}-AIC_{B}.$ For $\lvert\Delta{AIC}\rvert<2$, the models are
equally consistent with the data, when $2<\lvert\Delta{AIC}\rvert<6$, the
evidence is weak; while for $6<\lvert\Delta{AIC}\rvert<10$, the evidence is
strong. Finally, for $\lvert\Delta{AIC}\rvert>10$, there is a clear evidence
favoring for the model with the lower AIC.

Therefore, from the $\chi_{\min}^{2}$ parameters obtained before, we conclude
that for the datasets $D_{1}$ and $D_{4}$ there exist a weak evidence in favor
of the $\Lambda$CDM. Nevertheless, for the rest four combinations of data,
that is, datasets $D_{2}$,~$D_{3}$,~$D_{5}$ and $D_{6}$, the application of
Akaike's scale reveal that the two models are statistically indistinguishable.

On the other hand, as we discussed before, from Fig. \ref{fig7}, parameter
$\lambda$ has a small impact on the obtained physical solution.\ Hence, by
considering a fixed value, i.e. $\lambda=2$, from a model obtained from
theoretical constraints, we reduce the dimension of the parametric space by
one, thus, the AIC reveal that none of the datasets have a preferred model and
$f\left(  Q\right)  $-gravity fits the data similar with the $\Lambda$CDM. We
remark that we have verified this conclusion numerically.

\section{Conclusions}

\label{sec5}

We examined the noncoincidence $f\left(  Q\right)  $-gravity with
matter-gravity coupling as a geometric dark energy candidate for the
explanation of the late-time observational data. In our cosmological framework
of a spatially flat FLRW universe, we considered a connection which naturally
follows from the general scenario of a universe with nonzero spatial
curvature. This connection has a nontrivial contribution to the dark energy dynamics.

For the nonlinear function $f\left(  Q\right)  $ we select the simple
power-law function $f\left(  Q\right)  =f_{0}Q^{n}$, which introduces the
minimum number of free parameters in the cosmological dynamics. This can be
seen as an asymptotic limit of a more generic function $f\left(  Q\right)  $.
Moreover, for the matter source is assumed to be described by the Lagrangian
function of a dust fluid with a nonzero coupling function to gravity. The
presence of the coupling function, modify the continious equation and
introduces new dynamical behaviour for the gravitational model. We adopted the
scalar field description and wrote the gravitational field equations expressed
in terms of a multiscalar field theory, in this approach the theory belongs to
the family of matter-scalar field interaction within the Jordan frame, such
that when the modified field equations are expressed with the use of the
Einstein tensor, it is not necessary to introduce an effective time varying
gravitational coupling parameter.

We assume the power-law $f\left(  Q\right)  $ model, which leads to a
power-law scalar field potential. This choice is made in order to introduce
the minimum number of degrees of freedom in the gravitational field. We tested
this model by using SNIa data of the PP, U3 and DD catalogues, combined with
the OHD and BAO observations from DESI DR2. For six different combinations of
these data sets, we constrained the free parameters of the model and we found
consistent constraints. The index $n$ of the power-law theory is constrained
to have a finite value around $n\simeq2$. Nevertheless, the index $n$ has a
small impact on the evolution of the cosmological parameters.

\begin{figure}[t]
\centering\includegraphics[width=0.4\textwidth]{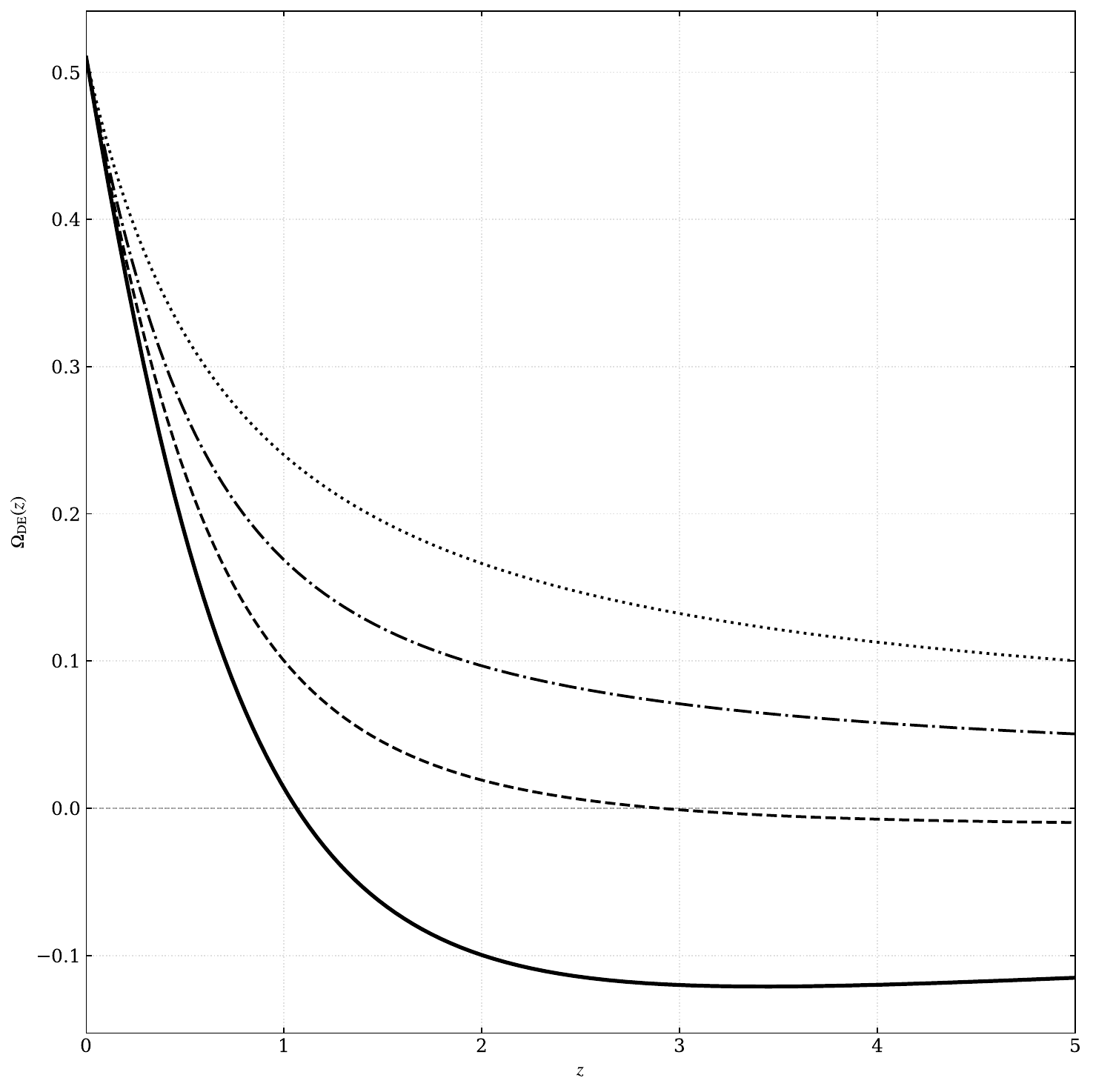}\caption{Dynamical
evolution of the $\Omega_{DE}(z)$ for large redshifts, where $\Omega_{DE}(z)$
change sign. The solid and the dashed lines are for initial conditions with
$x_{0}\xi_{0}<0$, while the dotted and dashed-dotted lies are for initial
conditions with $x_{0}\xi_{0}>0$.}%
\label{mat}%
\end{figure}

The numerical solutions for large redshifts reveal a phantom behaviour for the
geometric dark energy. That is, parameter $\Omega_{DE}\left(  z\right)  $ can
changes sign, as can be shown in Fig. \ref{mat}. Cosmological models with
change sign for the dark energy density have been examined before. For
instance, for the graduated dark energy model the effective cosmological
constant may have spontaneous sign switch during cosmic evolution
\cite{Akarsu:2019hmw}, while in quantum vacuum effects can allow such
transition as discussed in \cite{Anchordoqui:2023woo}. Such models have been
applied to address the cosmological tensions and the recent observation data
\cite{Toda:2024ncp,Akarsu:2025dmj}, for more details we refer the reader to
\cite{di2025cosmoverse}.  Thus, the equation of state parameter for the
effective dark energy component is defined in the phantom regime. That is the
reason that parameter $\Omega_{m0}$ has a larger value in comparison to
the~$\Lambda$CDM.

This phantom-like behaviour corresponds to the crossing of the null energy
condition, $\rho_{DE}+p_{DE}=0$, at the level of the dark energy component,
with $\rho_{DE}+p_{DE}<0$ characterising the phantom regime. It should be
noted, however, that the null energy condition strictly applies to the total
effective cosmic fluid rather than to individual components; apparent
violations at the component level may therefore reflect effective descriptions
of the underlying gravitational dynamics rather than a fundamental NEC
breakdown, for more details see \cite{Akarsu:2026anp,Gokcen:2026pkq}. Within
$f\left(  Q\right)  $-theory, the violation of this energy condition can be
easily explained when effective scalar fields are used to describe the
dynamical degrees of freedom \cite{Paliathanasis:2023pqp}. As discussed in
details in \cite{Basilakos:2025olm}, the dynamical degrees of freedom
introduced by the nonlinear function $f\left(  Q\right)  $, with the dynamics
provived by the noncoincidence connection can give rise to an effective
quintom-like cosmological behavior, which allows for such a transition
\cite{Cai:2009zp}. 

We conclude that for the observational tests we performed in this work, for
the $f\left(  Q\right)  =f_{0}Q^{n}$-gravity we obtained smaller values for
the $\chi_{\min}^{2}$ than those of the $\Lambda$CDM, while the application of
the AIC reveals that the two models are statistically indistinguishable when
the U3 and DD catalogues are applied, whereas the $\Lambda$CDM has a weak
preference from the combined data with the PP SNIa catalogue.

This is the first systematic study investigating the impact of the specific
connection defined in the noncoincident gauge. In contrast to the coincidence
connection, where late-time acceleration typically requires the introduction
of the cosmological constant or more complex $f(Q)$ forms, the present
noncoincident consideration achieves viable cosmological model, already at the
quadratic order. While in the literature the coincidence connection is often
selected for reasons of simplicity, our analysis demonstrates that the
noncoincident connection $\Gamma{^{C}}$ is not only mathematically consistent
with the flat limit of a curved universe but also observationally viable for
the description of late-time data. Taking into account that in anisotropic
cosmologies or compact-star models the coincidence connection does not always
reproduce the General Relativity limit, this indicates that the coincidence
property should perhaps be viewed as a specific case rather than a fundamental
requirement of symmetric teleparallelism.

As a next step, we will explore whether cosmological observations indicate a
preferred connection for the description of cosmic history and introduce
linear perturbations to examine whether the noncoincident theory can address
cosmological tension problems.

\begin{acknowledgments}
AP thanks the support of VRIDT through Resoluci\'{o}n VRIDT No. 096/2022 and
Resoluci\'{o}n VRIDT No. 021/2026. Part of this study was supported by
FONDECYT 1240514.
\end{acknowledgments}

\bibliographystyle{plain}
\bibliography{biblio}

\begin{thebibliography}{112}
\expandafter\ifx\csname natexlab\endcsname\relax\def\natexlab#1{#1}\fi
\expandafter\ifx\csname bibnamefont\endcsname\relax
  \def\bibnamefont#1{#1}\fi
\expandafter\ifx\csname bibfnamefont\endcsname\relax
  \def\bibfnamefont#1{#1}\fi
\expandafter\ifx\csname citenamefont\endcsname\relax
  \def\citenamefont#1{#1}\fi
\expandafter\ifx\csname url\endcsname\relax
  \def\url#1{\texttt{#1}}\fi
\expandafter\ifx\csname urlprefix\endcsname\relax\def\urlprefix{URL }\fi
\providecommand{\bibinfo}[2]{#2}
\providecommand{\eprint}[2][]{\url{#2}}

\bibitem[{\citenamefont{Riess et~al.}(1998)}]{SupernovaSearchTeam:1998fmf}
\bibinfo{author}{\bibfnamefont{A.~G.} \bibnamefont{Riess}} \bibnamefont{et~al.}
  (\bibinfo{collaboration}{Supernova Search Team}), \bibinfo{journal}{Astron.
  J.} \textbf{\bibinfo{volume}{116}}, \bibinfo{pages}{1009}
  (\bibinfo{year}{1998}), \eprint{astro-ph/9805201}.

\bibitem[{\citenamefont{Tegmark et~al.}(2004)}]{SDSS:2003tbn}
\bibinfo{author}{\bibfnamefont{M.}~\bibnamefont{Tegmark}} \bibnamefont{et~al.}
  (\bibinfo{collaboration}{SDSS}), \bibinfo{journal}{Astrophys. J.}
  \textbf{\bibinfo{volume}{606}}, \bibinfo{pages}{702} (\bibinfo{year}{2004}),
  \eprint{astro-ph/0310725}.

\bibitem[{\citenamefont{Kowalski
  et~al.}(2008)}]{SupernovaCosmologyProject:2008ojh}
\bibinfo{author}{\bibfnamefont{M.}~\bibnamefont{Kowalski}} \bibnamefont{et~al.}
  (\bibinfo{collaboration}{Supernova Cosmology Project}),
  \bibinfo{journal}{Astrophys. J.} \textbf{\bibinfo{volume}{686}},
  \bibinfo{pages}{749} (\bibinfo{year}{2008}), \eprint{0804.4142}.

\bibitem[{\citenamefont{Bret{\'o}n and Montiel}(2013)}]{Breton:2013twa}
\bibinfo{author}{\bibfnamefont{N.}~\bibnamefont{Bret{\'o}n}} \bibnamefont{and}
  \bibinfo{author}{\bibfnamefont{A.}~\bibnamefont{Montiel}},
  \bibinfo{journal}{Phys. Rev. D} \textbf{\bibinfo{volume}{87}},
  \bibinfo{pages}{063527} (\bibinfo{year}{2013}), \eprint{1303.1574}.

\bibitem[{\citenamefont{Frieman et~al.}(2008)\citenamefont{Frieman, Turner, and
  Huterer}}]{Frieman:2008sn}
\bibinfo{author}{\bibfnamefont{J.}~\bibnamefont{Frieman}},
  \bibinfo{author}{\bibfnamefont{M.}~\bibnamefont{Turner}}, \bibnamefont{and}
  \bibinfo{author}{\bibfnamefont{D.}~\bibnamefont{Huterer}},
  \bibinfo{journal}{Ann. Rev. Astron. Astrophys.}
  \textbf{\bibinfo{volume}{46}}, \bibinfo{pages}{385} (\bibinfo{year}{2008}),
  \eprint{0803.0982}.

\bibitem[{\citenamefont{de~Cruz~Perez et~al.}(2024)\citenamefont{de~Cruz~Perez,
  Park, and Ratra}}]{deCruzPerez:2024shj}
\bibinfo{author}{\bibfnamefont{J.}~\bibnamefont{de~Cruz~Perez}},
  \bibinfo{author}{\bibfnamefont{C.-G.} \bibnamefont{Park}}, \bibnamefont{and}
  \bibinfo{author}{\bibfnamefont{B.}~\bibnamefont{Ratra}},
  \bibinfo{journal}{Phys. Rev. D} \textbf{\bibinfo{volume}{110}},
  \bibinfo{pages}{023506} (\bibinfo{year}{2024}), \eprint{2404.19194}.

\bibitem[{\citenamefont{Park et~al.}(2024)\citenamefont{Park,
  de~Cruz~P{\'e}rez, and Ratra}}]{Park:2024vrw}
\bibinfo{author}{\bibfnamefont{C.-G.} \bibnamefont{Park}},
  \bibinfo{author}{\bibfnamefont{J.}~\bibnamefont{de~Cruz~P{\'e}rez}},
  \bibnamefont{and} \bibinfo{author}{\bibfnamefont{B.}~\bibnamefont{Ratra}},
  \bibinfo{journal}{Phys. Rev. D} \textbf{\bibinfo{volume}{110}},
  \bibinfo{pages}{123533} (\bibinfo{year}{2024}), \eprint{2405.00502}.

\bibitem[{\citenamefont{Bechtol et~al.}(2025)}]{DES:2025key}
\bibinfo{author}{\bibfnamefont{K.}~\bibnamefont{Bechtol}} \bibnamefont{et~al.}
  (\bibinfo{collaboration}{DES}) (\bibinfo{year}{2025}), \eprint{2501.05739}.

\bibitem[{\citenamefont{Lodha et~al.}(2025)}]{DESI:2025fii}
\bibinfo{author}{\bibfnamefont{K.}~\bibnamefont{Lodha}} \bibnamefont{et~al.}
  (\bibinfo{collaboration}{DESI}) (\bibinfo{year}{2025}), \eprint{2503.14743}.

\bibitem[{\citenamefont{Abdul~Karim et~al.}(2025{\natexlab{a}})}]{DESI:2025zgx}
\bibinfo{author}{\bibfnamefont{M.}~\bibnamefont{Abdul~Karim}}
  \bibnamefont{et~al.} (\bibinfo{collaboration}{DESI})
  (\bibinfo{year}{2025}{\natexlab{a}}), \eprint{2503.14738}.

\bibitem[{\citenamefont{Abdul~Karim et~al.}(2025{\natexlab{b}})}]{DESI:2025zpo}
\bibinfo{author}{\bibfnamefont{M.}~\bibnamefont{Abdul~Karim}}
  \bibnamefont{et~al.} (\bibinfo{collaboration}{DESI})
  (\bibinfo{year}{2025}{\natexlab{b}}), \eprint{2503.14739}.

\bibitem[{\citenamefont{Feng and Lu}(2011)}]{Feng:2011zzo}
\bibinfo{author}{\bibfnamefont{L.}~\bibnamefont{Feng}} \bibnamefont{and}
  \bibinfo{author}{\bibfnamefont{T.}~\bibnamefont{Lu}}, \bibinfo{journal}{JCAP}
  \textbf{\bibinfo{volume}{11}}, \bibinfo{pages}{034} (\bibinfo{year}{2011}),
  \eprint{1203.1784}.

\bibitem[{\citenamefont{Barboza and Alcaniz}(2008)}]{Barboza:2008rh}
\bibinfo{author}{\bibfnamefont{E.~M.} \bibnamefont{Barboza},
  \bibfnamefont{Jr.}} \bibnamefont{and} \bibinfo{author}{\bibfnamefont{J.~S.}
  \bibnamefont{Alcaniz}}, \bibinfo{journal}{Phys. Lett. B}
  \textbf{\bibinfo{volume}{666}}, \bibinfo{pages}{415} (\bibinfo{year}{2008}),
  \eprint{0805.1713}.

\bibitem[{\citenamefont{Bernardo et~al.}(2022)\citenamefont{Bernardo,
  Grand{\'o}n, Said~Levi, and C{\'a}rdenas}}]{Bernardo:2021cxi}
\bibinfo{author}{\bibfnamefont{R.~C.} \bibnamefont{Bernardo}},
  \bibinfo{author}{\bibfnamefont{D.}~\bibnamefont{Grand{\'o}n}},
  \bibinfo{author}{\bibfnamefont{J.}~\bibnamefont{Said~Levi}},
  \bibnamefont{and} \bibinfo{author}{\bibfnamefont{V.~H.}
  \bibnamefont{C{\'a}rdenas}}, \bibinfo{journal}{Phys. Dark Univ.}
  \textbf{\bibinfo{volume}{36}}, \bibinfo{pages}{101017}
  (\bibinfo{year}{2022}), \eprint{2111.08289}.

\bibitem[{\citenamefont{Rezaei}(2024)}]{Rezaei:2024vtg}
\bibinfo{author}{\bibfnamefont{M.}~\bibnamefont{Rezaei}},
  \bibinfo{journal}{Astrophys. J.} \textbf{\bibinfo{volume}{967}},
  \bibinfo{pages}{2} (\bibinfo{year}{2024}), \eprint{2403.18968}.

\bibitem[{\citenamefont{Escamilla et~al.}(2025)\citenamefont{Escamilla, Pan,
  Di~Valentino, Paliathanasis, V{\'a}zquez, and Yang}}]{Escamilla:2024fzq}
\bibinfo{author}{\bibfnamefont{L.~A.} \bibnamefont{Escamilla}},
  \bibinfo{author}{\bibfnamefont{S.}~\bibnamefont{Pan}},
  \bibinfo{author}{\bibfnamefont{E.}~\bibnamefont{Di~Valentino}},
  \bibinfo{author}{\bibfnamefont{A.}~\bibnamefont{Paliathanasis}},
  \bibinfo{author}{\bibfnamefont{J.~A.} \bibnamefont{V{\'a}zquez}},
  \bibnamefont{and} \bibinfo{author}{\bibfnamefont{W.}~\bibnamefont{Yang}},
  \bibinfo{journal}{Phys. Rev. D} \textbf{\bibinfo{volume}{111}},
  \bibinfo{pages}{023531} (\bibinfo{year}{2025}), \eprint{2404.00181}.

\bibitem[{\citenamefont{Hur et~al.}(2025)\citenamefont{Hur, Jejjala, Kavic,
  Minic, and Takeuchi}}]{Hur:2025lqc}
\bibinfo{author}{\bibfnamefont{S.}~\bibnamefont{Hur}},
  \bibinfo{author}{\bibfnamefont{V.}~\bibnamefont{Jejjala}},
  \bibinfo{author}{\bibfnamefont{M.~J.} \bibnamefont{Kavic}},
  \bibinfo{author}{\bibfnamefont{D.}~\bibnamefont{Minic}}, \bibnamefont{and}
  \bibinfo{author}{\bibfnamefont{T.}~\bibnamefont{Takeuchi}},
  \bibinfo{journal}{JCAP} \textbf{\bibinfo{volume}{11}}, \bibinfo{pages}{054}
  (\bibinfo{year}{2025}), \eprint{2503.20854}.

\bibitem[{\citenamefont{Chakraborty et~al.}(2025)\citenamefont{Chakraborty,
  Louw, Dunsby, MacDevette, and de~la Cruz~Dombriz}}]{Chakraborty:2025rvc}
\bibinfo{author}{\bibfnamefont{S.}~\bibnamefont{Chakraborty}},
  \bibinfo{author}{\bibfnamefont{C.}~\bibnamefont{Louw}},
  \bibinfo{author}{\bibfnamefont{P.~K.~S.} \bibnamefont{Dunsby}},
  \bibinfo{author}{\bibfnamefont{K.}~\bibnamefont{MacDevette}},
  \bibnamefont{and} \bibinfo{author}{\bibfnamefont{A.}~\bibnamefont{de~la
  Cruz~Dombriz}}, \bibinfo{journal}{Phys. Rev. D}
  \textbf{\bibinfo{volume}{112}}, \bibinfo{pages}{103518}
  (\bibinfo{year}{2025}), \eprint{2508.09813}.

\bibitem[{\citenamefont{Luongo and Muccino}(2025)}]{Luongo:2024zhc}
\bibinfo{author}{\bibfnamefont{O.}~\bibnamefont{Luongo}} \bibnamefont{and}
  \bibinfo{author}{\bibfnamefont{M.}~\bibnamefont{Muccino}},
  \bibinfo{journal}{Astron. Astrophys.} \textbf{\bibinfo{volume}{693}},
  \bibinfo{pages}{A187} (\bibinfo{year}{2025}), \eprint{2411.04901}.

\bibitem[{\citenamefont{Ratra and Peebles}(1988)}]{Ratra:1987rm}
\bibinfo{author}{\bibfnamefont{B.}~\bibnamefont{Ratra}} \bibnamefont{and}
  \bibinfo{author}{\bibfnamefont{P.~J.~E.} \bibnamefont{Peebles}},
  \bibinfo{journal}{Phys. Rev. D} \textbf{\bibinfo{volume}{37}},
  \bibinfo{pages}{3406} (\bibinfo{year}{1988}).

\bibitem[{\citenamefont{Tsujikawa}(2013)}]{Tsujikawa:2013fta}
\bibinfo{author}{\bibfnamefont{S.}~\bibnamefont{Tsujikawa}},
  \bibinfo{journal}{Class. Quant. Grav.} \textbf{\bibinfo{volume}{30}},
  \bibinfo{pages}{214003} (\bibinfo{year}{2013}), \eprint{1304.1961}.

\bibitem[{\citenamefont{Gorini et~al.}(2004)\citenamefont{Gorini, Kamenshchik,
  Moschella, and Pasquier}}]{Gorini:2004by}
\bibinfo{author}{\bibfnamefont{V.}~\bibnamefont{Gorini}},
  \bibinfo{author}{\bibfnamefont{A.}~\bibnamefont{Kamenshchik}},
  \bibinfo{author}{\bibfnamefont{U.}~\bibnamefont{Moschella}},
  \bibnamefont{and} \bibinfo{author}{\bibfnamefont{V.}~\bibnamefont{Pasquier}},
  in \emph{\bibinfo{booktitle}{{10th Marcel Grossmann Meeting on Recent
  Developments in Theoretical and Experimental General Relativity, Gravitation
  and Relativistic Field Theories (MG X MMIII)}}} (\bibinfo{year}{2004}), pp.
  \bibinfo{pages}{840--859}, \eprint{gr-qc/0403062}.

\bibitem[{\citenamefont{von Marttens et~al.}(2023)\citenamefont{von Marttens,
  Barbosa, and Alcaniz}}]{vonMarttens:2022xyr}
\bibinfo{author}{\bibfnamefont{R.}~\bibnamefont{von Marttens}},
  \bibinfo{author}{\bibfnamefont{D.}~\bibnamefont{Barbosa}}, \bibnamefont{and}
  \bibinfo{author}{\bibfnamefont{J.}~\bibnamefont{Alcaniz}},
  \bibinfo{journal}{JCAP} \textbf{\bibinfo{volume}{04}}, \bibinfo{pages}{052}
  (\bibinfo{year}{2023}), \eprint{2208.06302}.

\bibitem[{\citenamefont{de~Putter and Linder}(2007)}]{dePutter:2007ny}
\bibinfo{author}{\bibfnamefont{R.}~\bibnamefont{de~Putter}} \bibnamefont{and}
  \bibinfo{author}{\bibfnamefont{E.~V.} \bibnamefont{Linder}},
  \bibinfo{journal}{Astropart. Phys.} \textbf{\bibinfo{volume}{28}},
  \bibinfo{pages}{263} (\bibinfo{year}{2007}), \eprint{0705.0400}.

\bibitem[{\citenamefont{Bagla et~al.}(2003)\citenamefont{Bagla, Jassal, and
  Padmanabhan}}]{Bagla:2002yn}
\bibinfo{author}{\bibfnamefont{J.~S.} \bibnamefont{Bagla}},
  \bibinfo{author}{\bibfnamefont{H.~K.} \bibnamefont{Jassal}},
  \bibnamefont{and}
  \bibinfo{author}{\bibfnamefont{T.}~\bibnamefont{Padmanabhan}},
  \bibinfo{journal}{Phys. Rev. D} \textbf{\bibinfo{volume}{67}},
  \bibinfo{pages}{063504} (\bibinfo{year}{2003}), \eprint{astro-ph/0212198}.

\bibitem[{\citenamefont{Farnes}(2018)}]{Farnes:2017gbf}
\bibinfo{author}{\bibfnamefont{J.~S.} \bibnamefont{Farnes}},
  \bibinfo{journal}{Astron. Astrophys.} \textbf{\bibinfo{volume}{620}},
  \bibinfo{pages}{A92} (\bibinfo{year}{2018}), \eprint{1712.07962}.

\bibitem[{\citenamefont{Tsujikawa}(2008)}]{Tsujikawa:2007xu}
\bibinfo{author}{\bibfnamefont{S.}~\bibnamefont{Tsujikawa}},
  \bibinfo{journal}{Phys. Rev. D} \textbf{\bibinfo{volume}{77}},
  \bibinfo{pages}{023507} (\bibinfo{year}{2008}), \eprint{0709.1391}.

\bibitem[{\citenamefont{Nojiri et~al.}(2017)\citenamefont{Nojiri, Odintsov, and
  Oikonomou}}]{Nojiri:2017ncd}
\bibinfo{author}{\bibfnamefont{S.}~\bibnamefont{Nojiri}},
  \bibinfo{author}{\bibfnamefont{S.~D.} \bibnamefont{Odintsov}},
  \bibnamefont{and} \bibinfo{author}{\bibfnamefont{V.~K.}
  \bibnamefont{Oikonomou}}, \bibinfo{journal}{Phys. Rept.}
  \textbf{\bibinfo{volume}{692}}, \bibinfo{pages}{1} (\bibinfo{year}{2017}),
  \eprint{1705.11098}.

\bibitem[{\citenamefont{Bahamonde et~al.}(2023)\citenamefont{Bahamonde,
  Dialektopoulos, Escamilla-Rivera, Farrugia, Gakis, Hendry, Hohmann,
  Levi~Said, Mifsud, and Di~Valentino}}]{Bahamonde:2021gfp}
\bibinfo{author}{\bibfnamefont{S.}~\bibnamefont{Bahamonde}},
  \bibinfo{author}{\bibfnamefont{K.~F.} \bibnamefont{Dialektopoulos}},
  \bibinfo{author}{\bibfnamefont{C.}~\bibnamefont{Escamilla-Rivera}},
  \bibinfo{author}{\bibfnamefont{G.}~\bibnamefont{Farrugia}},
  \bibinfo{author}{\bibfnamefont{V.}~\bibnamefont{Gakis}},
  \bibinfo{author}{\bibfnamefont{M.}~\bibnamefont{Hendry}},
  \bibinfo{author}{\bibfnamefont{M.}~\bibnamefont{Hohmann}},
  \bibinfo{author}{\bibfnamefont{J.}~\bibnamefont{Levi~Said}},
  \bibinfo{author}{\bibfnamefont{J.}~\bibnamefont{Mifsud}}, \bibnamefont{and}
  \bibinfo{author}{\bibfnamefont{E.}~\bibnamefont{Di~Valentino}},
  \bibinfo{journal}{Rept. Prog. Phys.} \textbf{\bibinfo{volume}{86}},
  \bibinfo{pages}{026901} (\bibinfo{year}{2023}), \eprint{2106.13793}.

\bibitem[{\citenamefont{Nojiri and Odintsov}(2011)}]{Nojiri:2008nt}
\bibinfo{author}{\bibfnamefont{S.}~\bibnamefont{Nojiri}} \bibnamefont{and}
  \bibinfo{author}{\bibfnamefont{S.~D.} \bibnamefont{Odintsov}},
  \bibinfo{journal}{TSPU Bulletin} \textbf{\bibinfo{volume}{N8(110)}},
  \bibinfo{pages}{7} (\bibinfo{year}{2011}), \eprint{0807.0685}.

\bibitem[{\citenamefont{Heisenberg}(2019)}]{Heisenberg:2018vsk}
\bibinfo{author}{\bibfnamefont{L.}~\bibnamefont{Heisenberg}},
  \bibinfo{journal}{Phys. Rept.} \textbf{\bibinfo{volume}{796}},
  \bibinfo{pages}{1} (\bibinfo{year}{2019}), \eprint{1807.01725}.

\bibitem[{\citenamefont{Csillag and Jensko}(2025)}]{Csillag:2025gnz}
\bibinfo{author}{\bibfnamefont{L.}~\bibnamefont{Csillag}} \bibnamefont{and}
  \bibinfo{author}{\bibfnamefont{E.}~\bibnamefont{Jensko}}
  (\bibinfo{year}{2025}), \eprint{2505.15975}.

\bibitem[{\citenamefont{Chaudhary et~al.}(2024)\citenamefont{Chaudhary,
  Csillag, and Harko}}]{Chaudhary:2024jkj}
\bibinfo{author}{\bibfnamefont{H.}~\bibnamefont{Chaudhary}},
  \bibinfo{author}{\bibfnamefont{L.}~\bibnamefont{Csillag}}, \bibnamefont{and}
  \bibinfo{author}{\bibfnamefont{T.}~\bibnamefont{Harko}},
  \bibinfo{journal}{Universe} \textbf{\bibinfo{volume}{10}},
  \bibinfo{pages}{419} (\bibinfo{year}{2024}), \eprint{2411.03060}.

\bibitem[{\citenamefont{Beltr{\'a}n~Jim{\'e}nez
  et~al.}(2020)\citenamefont{Beltr{\'a}n~Jim{\'e}nez, Heisenberg, Koivisto, and
  Pekar}}]{BeltranJimenez:2019tme}
\bibinfo{author}{\bibfnamefont{J.}~\bibnamefont{Beltr{\'a}n~Jim{\'e}nez}},
  \bibinfo{author}{\bibfnamefont{L.}~\bibnamefont{Heisenberg}},
  \bibinfo{author}{\bibfnamefont{T.~S.} \bibnamefont{Koivisto}},
  \bibnamefont{and} \bibinfo{author}{\bibfnamefont{S.}~\bibnamefont{Pekar}},
  \bibinfo{journal}{Phys. Rev. D} \textbf{\bibinfo{volume}{101}},
  \bibinfo{pages}{103507} (\bibinfo{year}{2020}), \eprint{1906.10027}.

\bibitem[{\citenamefont{Khyllep et~al.}(2021)\citenamefont{Khyllep,
  Paliathanasis, and Dutta}}]{Khyllep:2021pcu}
\bibinfo{author}{\bibfnamefont{W.}~\bibnamefont{Khyllep}},
  \bibinfo{author}{\bibfnamefont{A.}~\bibnamefont{Paliathanasis}},
  \bibnamefont{and} \bibinfo{author}{\bibfnamefont{J.}~\bibnamefont{Dutta}},
  \bibinfo{journal}{Phys. Rev. D} \textbf{\bibinfo{volume}{103}},
  \bibinfo{pages}{103521} (\bibinfo{year}{2021}), \eprint{2103.08372}.

\bibitem[{\citenamefont{Zhao}(2024)}]{Zhao:2024kri}
\bibinfo{author}{\bibfnamefont{D.}~\bibnamefont{Zhao}}, \bibinfo{journal}{Phys.
  Rev. D} \textbf{\bibinfo{volume}{110}}, \bibinfo{pages}{124034}
  (\bibinfo{year}{2024}), \eprint{2404.16299}.

\bibitem[{\citenamefont{De et~al.}(2024)\citenamefont{De, Loo, and
  Saridakis}}]{De:2023xua}
\bibinfo{author}{\bibfnamefont{A.}~\bibnamefont{De}},
  \bibinfo{author}{\bibfnamefont{T.-H.} \bibnamefont{Loo}}, \bibnamefont{and}
  \bibinfo{author}{\bibfnamefont{E.~N.} \bibnamefont{Saridakis}},
  \bibinfo{journal}{JCAP} \textbf{\bibinfo{volume}{03}}, \bibinfo{pages}{050}
  (\bibinfo{year}{2024}), \eprint{2308.00652}.

\bibitem[{\citenamefont{Gakis et~al.}(2020)\citenamefont{Gakis,
  Kr{\v{s}}{\v{s}}{\'a}k, Levi~Said, and Saridakis}}]{Gakis:2019rdd}
\bibinfo{author}{\bibfnamefont{V.}~\bibnamefont{Gakis}},
  \bibinfo{author}{\bibfnamefont{M.}~\bibnamefont{Kr{\v{s}}{\v{s}}{\'a}k}},
  \bibinfo{author}{\bibfnamefont{J.}~\bibnamefont{Levi~Said}},
  \bibnamefont{and} \bibinfo{author}{\bibfnamefont{E.~N.}
  \bibnamefont{Saridakis}}, \bibinfo{journal}{Phys. Rev. D}
  \textbf{\bibinfo{volume}{101}}, \bibinfo{pages}{064024}
  (\bibinfo{year}{2020}), \eprint{1908.05741}.

\bibitem[{\citenamefont{Nojiri and
  Odintsov}(2024{\natexlab{a}})}]{Nojiri:2024zab}
\bibinfo{author}{\bibfnamefont{S.}~\bibnamefont{Nojiri}} \bibnamefont{and}
  \bibinfo{author}{\bibfnamefont{S.~D.} \bibnamefont{Odintsov}},
  \bibinfo{journal}{Phys. Dark Univ.} \textbf{\bibinfo{volume}{45}},
  \bibinfo{pages}{101538} (\bibinfo{year}{2024}{\natexlab{a}}),
  \eprint{2404.18427}.

\bibitem[{\citenamefont{Heisenberg}(2024)}]{Heisenberg:2023lru}
\bibinfo{author}{\bibfnamefont{L.}~\bibnamefont{Heisenberg}},
  \bibinfo{journal}{Phys. Rept.} \textbf{\bibinfo{volume}{1066}},
  \bibinfo{pages}{1} (\bibinfo{year}{2024}), \eprint{2309.15958}.

\bibitem[{\citenamefont{Nojiri and
  Odintsov}(2024{\natexlab{b}})}]{Nojiri:2024hau}
\bibinfo{author}{\bibfnamefont{S.}~\bibnamefont{Nojiri}} \bibnamefont{and}
  \bibinfo{author}{\bibfnamefont{S.~D.} \bibnamefont{Odintsov}},
  \bibinfo{journal}{Fortsch. Phys.} \textbf{\bibinfo{volume}{72}},
  \bibinfo{pages}{2400113} (\bibinfo{year}{2024}{\natexlab{b}}),
  \eprint{2406.12558}.

\bibitem[{\citenamefont{Carloni and Luongo}(2025)}]{Carloni:2024ybx}
\bibinfo{author}{\bibfnamefont{Y.}~\bibnamefont{Carloni}} \bibnamefont{and}
  \bibinfo{author}{\bibfnamefont{O.}~\bibnamefont{Luongo}},
  \bibinfo{journal}{Class. Quant. Grav.} \textbf{\bibinfo{volume}{42}},
  \bibinfo{pages}{075014} (\bibinfo{year}{2025}), \eprint{2410.10935}.

\bibitem[{\citenamefont{Carloni et~al.}(2025)\citenamefont{Carloni, Luongo, and
  Paliathanasis}}]{Carloni:2025kev}
\bibinfo{author}{\bibfnamefont{Y.}~\bibnamefont{Carloni}},
  \bibinfo{author}{\bibfnamefont{O.}~\bibnamefont{Luongo}}, \bibnamefont{and}
  \bibinfo{author}{\bibfnamefont{A.}~\bibnamefont{Paliathanasis}},
  \bibinfo{journal}{Fortsch. Phys.} \textbf{\bibinfo{volume}{73}},
  \bibinfo{pages}{e70021} (\bibinfo{year}{2025}), \eprint{2504.19245}.

\bibitem[{\citenamefont{Nester and Yo}(1999)}]{Nester:1998mp}
\bibinfo{author}{\bibfnamefont{J.~M.} \bibnamefont{Nester}} \bibnamefont{and}
  \bibinfo{author}{\bibfnamefont{H.-J.} \bibnamefont{Yo}},
  \bibinfo{journal}{Chin. J. Phys.} \textbf{\bibinfo{volume}{37}},
  \bibinfo{pages}{113} (\bibinfo{year}{1999}), \eprint{gr-qc/9809049}.

\bibitem[{\citenamefont{Conroy and Koivisto}(2018)}]{Conroy:2017yln}
\bibinfo{author}{\bibfnamefont{A.}~\bibnamefont{Conroy}} \bibnamefont{and}
  \bibinfo{author}{\bibfnamefont{T.}~\bibnamefont{Koivisto}},
  \bibinfo{journal}{Eur. Phys. J. C} \textbf{\bibinfo{volume}{78}},
  \bibinfo{pages}{923} (\bibinfo{year}{2018}), \eprint{1710.05708}.

\bibitem[{\citenamefont{Zhao}(2022)}]{Zhao:2021zab}
\bibinfo{author}{\bibfnamefont{D.}~\bibnamefont{Zhao}}, \bibinfo{journal}{Eur.
  Phys. J. C} \textbf{\bibinfo{volume}{82}}, \bibinfo{pages}{303}
  (\bibinfo{year}{2022}), \eprint{2104.02483}.

\bibitem[{\citenamefont{Hohmann}(2021)}]{Hohmann:2021ast}
\bibinfo{author}{\bibfnamefont{M.}~\bibnamefont{Hohmann}},
  \bibinfo{journal}{Phys. Rev. D} \textbf{\bibinfo{volume}{104}},
  \bibinfo{pages}{124077} (\bibinfo{year}{2021}), \eprint{2109.01525}.

\bibitem[{\citenamefont{Dimakis
  et~al.}(2022{\natexlab{a}})\citenamefont{Dimakis, Paliathanasis, Roumeliotis,
  and Christodoulakis}}]{Dimakis:2022rkd}
\bibinfo{author}{\bibfnamefont{N.}~\bibnamefont{Dimakis}},
  \bibinfo{author}{\bibfnamefont{A.}~\bibnamefont{Paliathanasis}},
  \bibinfo{author}{\bibfnamefont{M.}~\bibnamefont{Roumeliotis}},
  \bibnamefont{and}
  \bibinfo{author}{\bibfnamefont{T.}~\bibnamefont{Christodoulakis}},
  \bibinfo{journal}{Phys. Rev. D} \textbf{\bibinfo{volume}{106}},
  \bibinfo{pages}{043509} (\bibinfo{year}{2022}{\natexlab{a}}),
  \eprint{2205.04680}.

\bibitem[{\citenamefont{Ayuso et~al.}(2025)\citenamefont{Ayuso,
  Bouhmadi-L{\'o}pez, Chen, Chew, Dialektopoulos, and Ong}}]{Ayuso:2025vkc}
\bibinfo{author}{\bibfnamefont{I.}~\bibnamefont{Ayuso}},
  \bibinfo{author}{\bibfnamefont{M.}~\bibnamefont{Bouhmadi-L{\'o}pez}},
  \bibinfo{author}{\bibfnamefont{C.-Y.} \bibnamefont{Chen}},
  \bibinfo{author}{\bibfnamefont{X.~Y.} \bibnamefont{Chew}},
  \bibinfo{author}{\bibfnamefont{K.}~\bibnamefont{Dialektopoulos}},
  \bibnamefont{and} \bibinfo{author}{\bibfnamefont{Y.~C.} \bibnamefont{Ong}},
  \bibinfo{journal}{JCAP} \textbf{\bibinfo{volume}{11}}, \bibinfo{pages}{068}
  (\bibinfo{year}{2025}), \eprint{2506.03506}.

\bibitem[{\citenamefont{Dimakis et~al.}(2023)\citenamefont{Dimakis,
  Roumeliotis, Paliathanasis, and Christodoulakis}}]{Dimakis:2023uib}
\bibinfo{author}{\bibfnamefont{N.}~\bibnamefont{Dimakis}},
  \bibinfo{author}{\bibfnamefont{M.}~\bibnamefont{Roumeliotis}},
  \bibinfo{author}{\bibfnamefont{A.}~\bibnamefont{Paliathanasis}},
  \bibnamefont{and}
  \bibinfo{author}{\bibfnamefont{T.}~\bibnamefont{Christodoulakis}},
  \bibinfo{journal}{Eur. Phys. J. C} \textbf{\bibinfo{volume}{83}},
  \bibinfo{pages}{794} (\bibinfo{year}{2023}), \eprint{2304.04419}.

\bibitem[{\citenamefont{Murtaza
  et~al.}(2025{\natexlab{a}})\citenamefont{Murtaza, Chakraborty, and
  De}}]{Murtaza:2025gme}
\bibinfo{author}{\bibfnamefont{G.}~\bibnamefont{Murtaza}},
  \bibinfo{author}{\bibfnamefont{S.}~\bibnamefont{Chakraborty}},
  \bibnamefont{and} \bibinfo{author}{\bibfnamefont{A.}~\bibnamefont{De}},
  \bibinfo{journal}{JCAP} \textbf{\bibinfo{volume}{08}}, \bibinfo{pages}{093}
  (\bibinfo{year}{2025}{\natexlab{a}}), \eprint{2504.21757}.

\bibitem[{\citenamefont{Paliathanasis}(2024)}]{Paliathanasis:2024xpy}
\bibinfo{author}{\bibfnamefont{A.}~\bibnamefont{Paliathanasis}},
  \bibinfo{journal}{Phys. Dark Univ.} \textbf{\bibinfo{volume}{46}},
  \bibinfo{pages}{101585} (\bibinfo{year}{2024}), \eprint{2407.15069}.

\bibitem[{\citenamefont{Bahamonde and J{\"a}rv}(2022)}]{Bahamonde:2022zgj}
\bibinfo{author}{\bibfnamefont{S.}~\bibnamefont{Bahamonde}} \bibnamefont{and}
  \bibinfo{author}{\bibfnamefont{L.}~\bibnamefont{J{\"a}rv}},
  \bibinfo{journal}{Eur. Phys. J. C} \textbf{\bibinfo{volume}{82}},
  \bibinfo{pages}{963} (\bibinfo{year}{2022}), \eprint{2208.01872}.

\bibitem[{\citenamefont{D'Ambrosio et~al.}(2022)\citenamefont{D'Ambrosio, Fell,
  Heisenberg, and Kuhn}}]{DAmbrosio:2021zpm}
\bibinfo{author}{\bibfnamefont{F.}~\bibnamefont{D'Ambrosio}},
  \bibinfo{author}{\bibfnamefont{S.~D.~B.} \bibnamefont{Fell}},
  \bibinfo{author}{\bibfnamefont{L.}~\bibnamefont{Heisenberg}},
  \bibnamefont{and} \bibinfo{author}{\bibfnamefont{S.}~\bibnamefont{Kuhn}},
  \bibinfo{journal}{Phys. Rev. D} \textbf{\bibinfo{volume}{105}},
  \bibinfo{pages}{024042} (\bibinfo{year}{2022}), \eprint{2109.03174}.

\bibitem[{\citenamefont{Dimakis et~al.}(2025)\citenamefont{Dimakis, Terzis,
  Paliathanasis, and Christodoulakis}}]{Dimakis:2024fan}
\bibinfo{author}{\bibfnamefont{N.}~\bibnamefont{Dimakis}},
  \bibinfo{author}{\bibfnamefont{P.~A.} \bibnamefont{Terzis}},
  \bibinfo{author}{\bibfnamefont{A.}~\bibnamefont{Paliathanasis}},
  \bibnamefont{and}
  \bibinfo{author}{\bibfnamefont{T.}~\bibnamefont{Christodoulakis}},
  \bibinfo{journal}{JHEAp} \textbf{\bibinfo{volume}{45}}, \bibinfo{pages}{273}
  (\bibinfo{year}{2025}), \eprint{2410.04513}.

\bibitem[{\citenamefont{Paliathanasis et~al.}(2024)\citenamefont{Paliathanasis,
  Dimakis, and Christodoulakis}}]{Paliathanasis:2023pqp}
\bibinfo{author}{\bibfnamefont{A.}~\bibnamefont{Paliathanasis}},
  \bibinfo{author}{\bibfnamefont{N.}~\bibnamefont{Dimakis}}, \bibnamefont{and}
  \bibinfo{author}{\bibfnamefont{T.}~\bibnamefont{Christodoulakis}},
  \bibinfo{journal}{Phys. Dark Univ.} \textbf{\bibinfo{volume}{43}},
  \bibinfo{pages}{101410} (\bibinfo{year}{2024}), \eprint{2308.15207}.

\bibitem[{\citenamefont{Paliathanasis}(2023{\natexlab{a}})}]{Paliathanasis:2023nkb}
\bibinfo{author}{\bibfnamefont{A.}~\bibnamefont{Paliathanasis}},
  \bibinfo{journal}{Phys. Dark Univ.} \textbf{\bibinfo{volume}{41}},
  \bibinfo{pages}{101255} (\bibinfo{year}{2023}{\natexlab{a}}),
  \eprint{2304.04219}.

\bibitem[{\citenamefont{Paliathanasis}(2023{\natexlab{b}})}]{Paliathanasis:2023raj}
\bibinfo{author}{\bibfnamefont{A.}~\bibnamefont{Paliathanasis}},
  \bibinfo{journal}{Phys. Dark Univ.} \textbf{\bibinfo{volume}{42}},
  \bibinfo{pages}{101355} (\bibinfo{year}{2023}{\natexlab{b}}),
  \eprint{2310.04195}.

\bibitem[{\citenamefont{Lazkoz et~al.}(2019)\citenamefont{Lazkoz, Lobo,
  Ortiz-Ba{\~n}os, and Salzano}}]{Lazkoz:2019sjl}
\bibinfo{author}{\bibfnamefont{R.}~\bibnamefont{Lazkoz}},
  \bibinfo{author}{\bibfnamefont{F.~S.~N.} \bibnamefont{Lobo}},
  \bibinfo{author}{\bibfnamefont{M.}~\bibnamefont{Ortiz-Ba{\~n}os}},
  \bibnamefont{and} \bibinfo{author}{\bibfnamefont{V.}~\bibnamefont{Salzano}},
  \bibinfo{journal}{Phys. Rev. D} \textbf{\bibinfo{volume}{100}},
  \bibinfo{pages}{104027} (\bibinfo{year}{2019}), \eprint{1907.13219}.

\bibitem[{\citenamefont{Anagnostopoulos
  et~al.}(2021)\citenamefont{Anagnostopoulos, Basilakos, and
  Saridakis}}]{Anagnostopoulos:2021ydo}
\bibinfo{author}{\bibfnamefont{F.~K.} \bibnamefont{Anagnostopoulos}},
  \bibinfo{author}{\bibfnamefont{S.}~\bibnamefont{Basilakos}},
  \bibnamefont{and} \bibinfo{author}{\bibfnamefont{E.~N.}
  \bibnamefont{Saridakis}}, \bibinfo{journal}{Phys. Lett. B}
  \textbf{\bibinfo{volume}{822}}, \bibinfo{pages}{136634}
  (\bibinfo{year}{2021}), \eprint{2104.15123}.

\bibitem[{\citenamefont{Atayde and Frusciante}(2021)}]{Atayde:2021pgb}
\bibinfo{author}{\bibfnamefont{L.}~\bibnamefont{Atayde}} \bibnamefont{and}
  \bibinfo{author}{\bibfnamefont{N.}~\bibnamefont{Frusciante}},
  \bibinfo{journal}{Phys. Rev. D} \textbf{\bibinfo{volume}{104}},
  \bibinfo{pages}{064052} (\bibinfo{year}{2021}), \eprint{2108.10832}.

\bibitem[{\citenamefont{Boiza et~al.}(2025)\citenamefont{Boiza, Petronikolou,
  Bouhmadi-L{\'o}pez, and Saridakis}}]{Boiza:2025xpn}
\bibinfo{author}{\bibfnamefont{C.~G.} \bibnamefont{Boiza}},
  \bibinfo{author}{\bibfnamefont{M.}~\bibnamefont{Petronikolou}},
  \bibinfo{author}{\bibfnamefont{M.}~\bibnamefont{Bouhmadi-L{\'o}pez}},
  \bibnamefont{and} \bibinfo{author}{\bibfnamefont{E.~N.}
  \bibnamefont{Saridakis}}, \bibinfo{journal}{JCAP}
  \textbf{\bibinfo{volume}{12}}, \bibinfo{pages}{011} (\bibinfo{year}{2025}),
  \eprint{2505.18264}.

\bibitem[{\citenamefont{Ferreira et~al.}(2023)\citenamefont{Ferreira, Barreiro,
  Mimoso, and Nunes}}]{Ferreira:2023awf}
\bibinfo{author}{\bibfnamefont{J.}~\bibnamefont{Ferreira}},
  \bibinfo{author}{\bibfnamefont{T.}~\bibnamefont{Barreiro}},
  \bibinfo{author}{\bibfnamefont{J.~P.} \bibnamefont{Mimoso}},
  \bibnamefont{and} \bibinfo{author}{\bibfnamefont{N.~J.} \bibnamefont{Nunes}},
  \bibinfo{journal}{Phys. Rev. D} \textbf{\bibinfo{volume}{108}},
  \bibinfo{pages}{063521} (\bibinfo{year}{2023}), \eprint{2306.10176}.

\bibitem[{\citenamefont{Shi}(2023)}]{Shi:2023kvu}
\bibinfo{author}{\bibfnamefont{J.}~\bibnamefont{Shi}}, \bibinfo{journal}{Eur.
  Phys. J. C} \textbf{\bibinfo{volume}{83}}, \bibinfo{pages}{951}
  (\bibinfo{year}{2023}), \eprint{2307.08103}.

\bibitem[{\citenamefont{Paliathanasis}(2025)}]{Paliathanasis:2025hjw}
\bibinfo{author}{\bibfnamefont{A.}~\bibnamefont{Paliathanasis}},
  \bibinfo{journal}{Phys. Dark Univ.} \textbf{\bibinfo{volume}{49}},
  \bibinfo{pages}{101993} (\bibinfo{year}{2025}), \eprint{2504.11132}.

\bibitem[{\citenamefont{Murtaza
  et~al.}(2025{\natexlab{b}})\citenamefont{Murtaza, De, and
  Paliathanasis}}]{Murtaza:2025klz}
\bibinfo{author}{\bibfnamefont{G.}~\bibnamefont{Murtaza}},
  \bibinfo{author}{\bibfnamefont{A.}~\bibnamefont{De}}, \bibnamefont{and}
  \bibinfo{author}{\bibfnamefont{A.}~\bibnamefont{Paliathanasis}}
  (\bibinfo{year}{2025}{\natexlab{b}}), \eprint{2509.00569}.

\bibitem[{\citenamefont{Abebe et~al.}(2025)\citenamefont{Abebe, Apostolopoulos,
  Giacomini, Leon, Moncada, and Paliathanasis}}]{Abebe:2025wos}
\bibinfo{author}{\bibfnamefont{A.}~\bibnamefont{Abebe}},
  \bibinfo{author}{\bibfnamefont{P.~S.} \bibnamefont{Apostolopoulos}},
  \bibinfo{author}{\bibfnamefont{A.}~\bibnamefont{Giacomini}},
  \bibinfo{author}{\bibfnamefont{G.}~\bibnamefont{Leon}},
  \bibinfo{author}{\bibfnamefont{F.}~\bibnamefont{Moncada}}, \bibnamefont{and}
  \bibinfo{author}{\bibfnamefont{A.}~\bibnamefont{Paliathanasis}}
  (\bibinfo{year}{2025}), \eprint{2510.00535}.

\bibitem[{\citenamefont{Gomes et~al.}(2024)\citenamefont{Gomes,
  Beltr{\'a}n~Jim{\'e}nez, Cano, and Koivisto}}]{Gomes:2023tur}
\bibinfo{author}{\bibfnamefont{D.~A.} \bibnamefont{Gomes}},
  \bibinfo{author}{\bibfnamefont{J.}~\bibnamefont{Beltr{\'a}n~Jim{\'e}nez}},
  \bibinfo{author}{\bibfnamefont{A.~J.} \bibnamefont{Cano}}, \bibnamefont{and}
  \bibinfo{author}{\bibfnamefont{T.~S.} \bibnamefont{Koivisto}},
  \bibinfo{journal}{Phys. Rev. Lett.} \textbf{\bibinfo{volume}{132}},
  \bibinfo{pages}{141401} (\bibinfo{year}{2024}), \eprint{2311.04201}.

\bibitem[{\citenamefont{Wang et~al.}(2016)\citenamefont{Wang, Abdalla,
  Atrio-Barandela, and Pavon}}]{Wang:2016lxa}
\bibinfo{author}{\bibfnamefont{B.}~\bibnamefont{Wang}},
  \bibinfo{author}{\bibfnamefont{E.}~\bibnamefont{Abdalla}},
  \bibinfo{author}{\bibfnamefont{F.}~\bibnamefont{Atrio-Barandela}},
  \bibnamefont{and} \bibinfo{author}{\bibfnamefont{D.}~\bibnamefont{Pavon}},
  \bibinfo{journal}{Rept. Prog. Phys.} \textbf{\bibinfo{volume}{79}},
  \bibinfo{pages}{096901} (\bibinfo{year}{2016}), \eprint{1603.08299}.

\bibitem[{\citenamefont{Lucca}(2021)}]{Lucca:2021dxo}
\bibinfo{author}{\bibfnamefont{M.}~\bibnamefont{Lucca}},
  \bibinfo{journal}{Phys. Dark Univ.} \textbf{\bibinfo{volume}{34}},
  \bibinfo{pages}{100899} (\bibinfo{year}{2021}), \eprint{2105.09249}.

\bibitem[{\citenamefont{Figueruelo et~al.}(2026)\citenamefont{Figueruelo,
  van~der Westhuizen, Abebe, and Di~Valentino}}]{Figueruelo:2026eis}
\bibinfo{author}{\bibfnamefont{D.}~\bibnamefont{Figueruelo}},
  \bibinfo{author}{\bibfnamefont{M.}~\bibnamefont{van~der Westhuizen}},
  \bibinfo{author}{\bibfnamefont{A.}~\bibnamefont{Abebe}}, \bibnamefont{and}
  \bibinfo{author}{\bibfnamefont{E.}~\bibnamefont{Di~Valentino}}
  (\bibinfo{year}{2026}), \eprint{2601.03122}.

\bibitem[{\citenamefont{Yang et~al.}(2019{\natexlab{a}})\citenamefont{Yang,
  Banerjee, Paliathanasis, and Pan}}]{Yang:2018qec}
\bibinfo{author}{\bibfnamefont{W.}~\bibnamefont{Yang}},
  \bibinfo{author}{\bibfnamefont{N.}~\bibnamefont{Banerjee}},
  \bibinfo{author}{\bibfnamefont{A.}~\bibnamefont{Paliathanasis}},
  \bibnamefont{and} \bibinfo{author}{\bibfnamefont{S.}~\bibnamefont{Pan}},
  \bibinfo{journal}{Phys. Dark Univ.} \textbf{\bibinfo{volume}{26}},
  \bibinfo{pages}{100383} (\bibinfo{year}{2019}{\natexlab{a}}),
  \eprint{1812.06854}.

\bibitem[{\citenamefont{Yang et~al.}(2019{\natexlab{b}})\citenamefont{Yang,
  Vagnozzi, Di~Valentino, Nunes, Pan, and Mota}}]{Yang:2019vni}
\bibinfo{author}{\bibfnamefont{W.}~\bibnamefont{Yang}},
  \bibinfo{author}{\bibfnamefont{S.}~\bibnamefont{Vagnozzi}},
  \bibinfo{author}{\bibfnamefont{E.}~\bibnamefont{Di~Valentino}},
  \bibinfo{author}{\bibfnamefont{R.~C.} \bibnamefont{Nunes}},
  \bibinfo{author}{\bibfnamefont{S.}~\bibnamefont{Pan}}, \bibnamefont{and}
  \bibinfo{author}{\bibfnamefont{D.~F.} \bibnamefont{Mota}},
  \bibinfo{journal}{JCAP} \textbf{\bibinfo{volume}{07}}, \bibinfo{pages}{037}
  (\bibinfo{year}{2019}{\natexlab{b}}), \eprint{1905.08286}.

\bibitem[{\citenamefont{Pan et~al.}(2020)\citenamefont{Pan, Sharov, and
  Yang}}]{Pan:2020zza}
\bibinfo{author}{\bibfnamefont{S.}~\bibnamefont{Pan}},
  \bibinfo{author}{\bibfnamefont{G.~S.} \bibnamefont{Sharov}},
  \bibnamefont{and} \bibinfo{author}{\bibfnamefont{W.}~\bibnamefont{Yang}},
  \bibinfo{journal}{Phys. Rev. D} \textbf{\bibinfo{volume}{101}},
  \bibinfo{pages}{103533} (\bibinfo{year}{2020}), \eprint{2001.03120}.

\bibitem[{\citenamefont{Benisty et~al.}(2024)\citenamefont{Benisty, Pan,
  Staicova, Di~Valentino, and Nunes}}]{Benisty:2024lmj}
\bibinfo{author}{\bibfnamefont{D.}~\bibnamefont{Benisty}},
  \bibinfo{author}{\bibfnamefont{S.}~\bibnamefont{Pan}},
  \bibinfo{author}{\bibfnamefont{D.}~\bibnamefont{Staicova}},
  \bibinfo{author}{\bibfnamefont{E.}~\bibnamefont{Di~Valentino}},
  \bibnamefont{and} \bibinfo{author}{\bibfnamefont{R.~C.} \bibnamefont{Nunes}},
  \bibinfo{journal}{Astron. Astrophys.} \textbf{\bibinfo{volume}{688}},
  \bibinfo{pages}{A156} (\bibinfo{year}{2024}), \eprint{2403.00056}.

\bibitem[{\citenamefont{Aboubrahim and Nath}(2026)}]{Aboubrahim:2026tks}
\bibinfo{author}{\bibfnamefont{A.}~\bibnamefont{Aboubrahim}} \bibnamefont{and}
  \bibinfo{author}{\bibfnamefont{P.}~\bibnamefont{Nath}}
  (\bibinfo{year}{2026}), \eprint{2601.19662}.

\bibitem[{\citenamefont{Guedezounme et~al.}(2025)\citenamefont{Guedezounme,
  Dinda, and Maartens}}]{Guedezounme:2025wav}
\bibinfo{author}{\bibfnamefont{S.~L.} \bibnamefont{Guedezounme}},
  \bibinfo{author}{\bibfnamefont{B.~R.} \bibnamefont{Dinda}}, \bibnamefont{and}
  \bibinfo{author}{\bibfnamefont{R.}~\bibnamefont{Maartens}}
  (\bibinfo{year}{2025}), \eprint{2507.18274}.

\bibitem[{\citenamefont{Zhai et~al.}(2025)\citenamefont{Zhai, de~Cesare, van~de
  Bruck, Di~Valentino, and Wilson-Ewing}}]{Zhai:2025hfi}
\bibinfo{author}{\bibfnamefont{Y.}~\bibnamefont{Zhai}},
  \bibinfo{author}{\bibfnamefont{M.}~\bibnamefont{de~Cesare}},
  \bibinfo{author}{\bibfnamefont{C.}~\bibnamefont{van~de Bruck}},
  \bibinfo{author}{\bibfnamefont{E.}~\bibnamefont{Di~Valentino}},
  \bibnamefont{and}
  \bibinfo{author}{\bibfnamefont{E.}~\bibnamefont{Wilson-Ewing}},
  \bibinfo{journal}{JCAP} \textbf{\bibinfo{volume}{11}}, \bibinfo{pages}{010}
  (\bibinfo{year}{2025}), \eprint{2503.15659}.

\bibitem[{\citenamefont{Paliathanasis}(2026)}]{Paliathanasis:2026ymi}
\bibinfo{author}{\bibfnamefont{A.}~\bibnamefont{Paliathanasis}}
  (\bibinfo{year}{2026}), \eprint{2601.02789}.

\bibitem[{\citenamefont{Li et~al.}(2026)\citenamefont{Li, Giar{\`e}, Du, Li,
  Di~Valentino, Zhang, and Zhang}}]{Li:2026xaz}
\bibinfo{author}{\bibfnamefont{T.-N.} \bibnamefont{Li}},
  \bibinfo{author}{\bibfnamefont{W.}~\bibnamefont{Giar{\`e}}},
  \bibinfo{author}{\bibfnamefont{G.-H.} \bibnamefont{Du}},
  \bibinfo{author}{\bibfnamefont{Y.-H.} \bibnamefont{Li}},
  \bibinfo{author}{\bibfnamefont{E.}~\bibnamefont{Di~Valentino}},
  \bibinfo{author}{\bibfnamefont{J.-F.} \bibnamefont{Zhang}}, \bibnamefont{and}
  \bibinfo{author}{\bibfnamefont{X.}~\bibnamefont{Zhang}}
  (\bibinfo{year}{2026}), \eprint{2601.07361}.

\bibitem[{\citenamefont{Salim and Sautu}(1996)}]{Salim:1996ei}
\bibinfo{author}{\bibfnamefont{J.~M.} \bibnamefont{Salim}} \bibnamefont{and}
  \bibinfo{author}{\bibfnamefont{S.~L.} \bibnamefont{Sautu}},
  \bibinfo{journal}{Class. Quant. Grav.} \textbf{\bibinfo{volume}{13}},
  \bibinfo{pages}{353} (\bibinfo{year}{1996}).

\bibitem[{\citenamefont{Romero et~al.}(2012)\citenamefont{Romero, Fonseca-Neto,
  and Pucheu}}]{Romero:2012hs}
\bibinfo{author}{\bibfnamefont{C.}~\bibnamefont{Romero}},
  \bibinfo{author}{\bibfnamefont{J.~B.} \bibnamefont{Fonseca-Neto}},
  \bibnamefont{and} \bibinfo{author}{\bibfnamefont{M.~L.}
  \bibnamefont{Pucheu}}, \bibinfo{journal}{Class. Quant. Grav.}
  \textbf{\bibinfo{volume}{29}}, \bibinfo{pages}{155015}
  (\bibinfo{year}{2012}), \eprint{1201.1469}.

\bibitem[{\citenamefont{Tsamparlis et~al.}(2013)\citenamefont{Tsamparlis,
  Paliathanasis, Basilakos, and Capozziello}}]{Tsamparlis:2013aza}
\bibinfo{author}{\bibfnamefont{M.}~\bibnamefont{Tsamparlis}},
  \bibinfo{author}{\bibfnamefont{A.}~\bibnamefont{Paliathanasis}},
  \bibinfo{author}{\bibfnamefont{S.}~\bibnamefont{Basilakos}},
  \bibnamefont{and}
  \bibinfo{author}{\bibfnamefont{S.}~\bibnamefont{Capozziello}},
  \bibinfo{journal}{Gen. Rel. Grav.} \textbf{\bibinfo{volume}{45}},
  \bibinfo{pages}{2003} (\bibinfo{year}{2013}), \eprint{1307.6694}.

\bibitem[{\citenamefont{Khoury and Weltman}(2004)}]{Khoury:2003rn}
\bibinfo{author}{\bibfnamefont{J.}~\bibnamefont{Khoury}} \bibnamefont{and}
  \bibinfo{author}{\bibfnamefont{A.}~\bibnamefont{Weltman}},
  \bibinfo{journal}{Phys. Rev. D} \textbf{\bibinfo{volume}{69}},
  \bibinfo{pages}{044026} (\bibinfo{year}{2004}), \eprint{astro-ph/0309411}.

\bibitem[{\citenamefont{Hinterbichler et~al.}(2011)\citenamefont{Hinterbichler,
  Khoury, Levy, and Matas}}]{Hinterbichler:2011ca}
\bibinfo{author}{\bibfnamefont{K.}~\bibnamefont{Hinterbichler}},
  \bibinfo{author}{\bibfnamefont{J.}~\bibnamefont{Khoury}},
  \bibinfo{author}{\bibfnamefont{A.}~\bibnamefont{Levy}}, \bibnamefont{and}
  \bibinfo{author}{\bibfnamefont{A.}~\bibnamefont{Matas}},
  \bibinfo{journal}{Phys. Rev. D} \textbf{\bibinfo{volume}{84}},
  \bibinfo{pages}{103521} (\bibinfo{year}{2011}), \eprint{1107.2112}.

\bibitem[{\citenamefont{Paliathanasis}(2023{\natexlab{c}})}]{Paliathanasis:2023moe}
\bibinfo{author}{\bibfnamefont{A.}~\bibnamefont{Paliathanasis}},
  \bibinfo{journal}{Eur. Phys. J. C} \textbf{\bibinfo{volume}{83}},
  \bibinfo{pages}{756} (\bibinfo{year}{2023}{\natexlab{c}}),
  \eprint{2308.10460}.

\bibitem[{\citenamefont{Luongo et~al.}(2026)\citenamefont{Luongo, Mengoni, and
  S{\'a}}}]{Luongo:2025ovo}
\bibinfo{author}{\bibfnamefont{O.}~\bibnamefont{Luongo}},
  \bibinfo{author}{\bibfnamefont{T.}~\bibnamefont{Mengoni}}, \bibnamefont{and}
  \bibinfo{author}{\bibfnamefont{P.~M.} \bibnamefont{S{\'a}}},
  \bibinfo{journal}{Phys. Rev. D} \textbf{\bibinfo{volume}{113}},
  \bibinfo{pages}{023541} (\bibinfo{year}{2026}), \eprint{2509.21200}.

\bibitem[{\citenamefont{Zhai et~al.}(2023)\citenamefont{Zhai, Giar{\`e}, van~de
  Bruck, Di~Valentino, Mena, and Nunes}}]{Zhai:2023yny}
\bibinfo{author}{\bibfnamefont{Y.}~\bibnamefont{Zhai}},
  \bibinfo{author}{\bibfnamefont{W.}~\bibnamefont{Giar{\`e}}},
  \bibinfo{author}{\bibfnamefont{C.}~\bibnamefont{van~de Bruck}},
  \bibinfo{author}{\bibfnamefont{E.}~\bibnamefont{Di~Valentino}},
  \bibinfo{author}{\bibfnamefont{O.}~\bibnamefont{Mena}}, \bibnamefont{and}
  \bibinfo{author}{\bibfnamefont{R.~C.} \bibnamefont{Nunes}},
  \bibinfo{journal}{JCAP} \textbf{\bibinfo{volume}{07}}, \bibinfo{pages}{032}
  (\bibinfo{year}{2023}), \eprint{2303.08201}.

\bibitem[{\citenamefont{Bertolami et~al.}(2007)\citenamefont{Bertolami,
  Boehmer, Harko, and Lobo}}]{Bertolami:2007gv}
\bibinfo{author}{\bibfnamefont{O.}~\bibnamefont{Bertolami}},
  \bibinfo{author}{\bibfnamefont{C.~G.} \bibnamefont{Boehmer}},
  \bibinfo{author}{\bibfnamefont{T.}~\bibnamefont{Harko}}, \bibnamefont{and}
  \bibinfo{author}{\bibfnamefont{F.~S.~N.} \bibnamefont{Lobo}},
  \bibinfo{journal}{Phys. Rev. D} \textbf{\bibinfo{volume}{75}},
  \bibinfo{pages}{104016} (\bibinfo{year}{2007}), \eprint{0704.1733}.

\bibitem[{\citenamefont{Bertolami et~al.}(2008)\citenamefont{Bertolami, Lobo,
  and Paramos}}]{Bertolami:2008ab}
\bibinfo{author}{\bibfnamefont{O.}~\bibnamefont{Bertolami}},
  \bibinfo{author}{\bibfnamefont{F.~S.~N.} \bibnamefont{Lobo}},
  \bibnamefont{and} \bibinfo{author}{\bibfnamefont{J.}~\bibnamefont{Paramos}},
  \bibinfo{journal}{Phys. Rev. D} \textbf{\bibinfo{volume}{78}},
  \bibinfo{pages}{064036} (\bibinfo{year}{2008}), \eprint{0806.4434}.

\bibitem[{\citenamefont{Haghani and Harko}(2021)}]{Haghani:2021fpx}
\bibinfo{author}{\bibfnamefont{Z.}~\bibnamefont{Haghani}} \bibnamefont{and}
  \bibinfo{author}{\bibfnamefont{T.}~\bibnamefont{Harko}},
  \bibinfo{journal}{Eur. Phys. J. C} \textbf{\bibinfo{volume}{81}},
  \bibinfo{pages}{615} (\bibinfo{year}{2021}), \eprint{2106.10644}.

\bibitem[{\citenamefont{Brout et~al.}(2022)}]{Brout:2022vxf}
\bibinfo{author}{\bibfnamefont{D.}~\bibnamefont{Brout}} \bibnamefont{et~al.},
  \bibinfo{journal}{Astrophys. J.} \textbf{\bibinfo{volume}{938}},
  \bibinfo{pages}{110} (\bibinfo{year}{2022}), \eprint{2202.04077}.

\bibitem[{\citenamefont{Rubin et~al.}(2023)}]{rubin2023union}
\bibinfo{author}{\bibfnamefont{D.}~\bibnamefont{Rubin}} \bibnamefont{et~al.},
  \bibinfo{journal}{arXiv preprint arXiv:2311.12098}  (\bibinfo{year}{2023}).

\bibitem[{\citenamefont{Popovic et~al.}(2025)}]{DES:2025sig}
\bibinfo{author}{\bibfnamefont{B.}~\bibnamefont{Popovic}} \bibnamefont{et~al.}
  (\bibinfo{collaboration}{DES}) (\bibinfo{year}{2025}), \eprint{2511.07517}.

\bibitem[{\citenamefont{Moresco et~al.}(2020)}]{moresco2020setting}
\bibinfo{author}{\bibfnamefont{M.}~\bibnamefont{Moresco}} \bibnamefont{et~al.},
  \bibinfo{journal}{The Astrophysical Journal} \textbf{\bibinfo{volume}{898}},
  \bibinfo{pages}{82} (\bibinfo{year}{2020}).

\bibitem[{\citenamefont{Loubser}(2025)}]{Loubser:2025fzl}
\bibinfo{author}{\bibfnamefont{S.~I.} \bibnamefont{Loubser}}
  (\bibinfo{year}{2025}), \eprint{2511.02730}.

\bibitem[{\citenamefont{Torrado and Lewis}(2019)}]{cob1}
\bibinfo{author}{\bibfnamefont{J.}~\bibnamefont{Torrado}} \bibnamefont{and}
  \bibinfo{author}{\bibfnamefont{A.}~\bibnamefont{Lewis}},
  \bibinfo{journal}{Astrophysics Source Code Library,}  (\bibinfo{year}{2019}),
  \eprint{ascl:1910.019}.

\bibitem[{\citenamefont{Torrado and Lewis}(2021)}]{cob2}
\bibinfo{author}{\bibfnamefont{J.}~\bibnamefont{Torrado}} \bibnamefont{and}
  \bibinfo{author}{\bibfnamefont{A.}~\bibnamefont{Lewis}},
  \bibinfo{journal}{JCAP} \textbf{\bibinfo{volume}{05}}, \bibinfo{pages}{057}
  (\bibinfo{year}{2021}), \eprint{2005.05290}.

\bibitem[{\citenamefont{Lewis and Bridle}(2002)}]{mcmc1}
\bibinfo{author}{\bibfnamefont{A.}~\bibnamefont{Lewis}} \bibnamefont{and}
  \bibinfo{author}{\bibfnamefont{S.}~\bibnamefont{Bridle}},
  \bibinfo{journal}{Phys. Rev. D} \textbf{\bibinfo{volume}{66}},
  \bibinfo{pages}{103511} (\bibinfo{year}{2002}), \eprint{astro-ph/0205436}.

\bibitem[{\citenamefont{Lewis}(2013)}]{mcmc2}
\bibinfo{author}{\bibfnamefont{A.}~\bibnamefont{Lewis}},
  \bibinfo{journal}{Phys. Rev. D} \textbf{\bibinfo{volume}{87}},
  \bibinfo{pages}{103529} (\bibinfo{year}{2013}), \eprint{1304.4473}.

\bibitem[{\citenamefont{Lewis}(2025)}]{getd}
\bibinfo{author}{\bibfnamefont{A.}~\bibnamefont{Lewis}},
  \bibinfo{journal}{JCAP} \textbf{\bibinfo{volume}{08}}, \bibinfo{pages}{025}
  (\bibinfo{year}{2025}), \eprint{1910.13970}.

\bibitem[{\citenamefont{Dimakis
  et~al.}(2022{\natexlab{b}})\citenamefont{Dimakis, Roumeliotis, Paliathanasis,
  Apostolopoulos, and Christodoulakis}}]{Dimakis:2022wkj}
\bibinfo{author}{\bibfnamefont{N.}~\bibnamefont{Dimakis}},
  \bibinfo{author}{\bibfnamefont{M.}~\bibnamefont{Roumeliotis}},
  \bibinfo{author}{\bibfnamefont{A.}~\bibnamefont{Paliathanasis}},
  \bibinfo{author}{\bibfnamefont{P.~S.} \bibnamefont{Apostolopoulos}},
  \bibnamefont{and}
  \bibinfo{author}{\bibfnamefont{T.}~\bibnamefont{Christodoulakis}},
  \bibinfo{journal}{Phys. Rev. D} \textbf{\bibinfo{volume}{106}},
  \bibinfo{pages}{123516} (\bibinfo{year}{2022}{\natexlab{b}}),
  \eprint{2210.10295}.

\bibitem[{\citenamefont{Akaike}(1974)}]{AIC}
\bibinfo{author}{\bibfnamefont{H.}~\bibnamefont{Akaike}},
  \bibinfo{journal}{IEEE Trans. Automatic Control}
  \textbf{\bibinfo{volume}{19}}, \bibinfo{pages}{716} (\bibinfo{year}{1974}).

\bibitem[{\citenamefont{Akarsu et~al.}(2020)\citenamefont{Akarsu, Barrow,
  Escamilla, and Vazquez}}]{Akarsu:2019hmw}
\bibinfo{author}{\bibfnamefont{{\"O}.}~\bibnamefont{Akarsu}},
  \bibinfo{author}{\bibfnamefont{J.~D.} \bibnamefont{Barrow}},
  \bibinfo{author}{\bibfnamefont{L.~A.} \bibnamefont{Escamilla}},
  \bibnamefont{and} \bibinfo{author}{\bibfnamefont{J.~A.}
  \bibnamefont{Vazquez}}, \bibinfo{journal}{Phys. Rev. D}
  \textbf{\bibinfo{volume}{101}}, \bibinfo{pages}{063528}
  (\bibinfo{year}{2020}), \eprint{1912.08751}.

\bibitem[{\citenamefont{Anchordoqui et~al.}(2024)\citenamefont{Anchordoqui,
  Antoniadis, and Lust}}]{Anchordoqui:2023woo}
\bibinfo{author}{\bibfnamefont{L.~A.} \bibnamefont{Anchordoqui}},
  \bibinfo{author}{\bibfnamefont{I.}~\bibnamefont{Antoniadis}},
  \bibnamefont{and} \bibinfo{author}{\bibfnamefont{D.}~\bibnamefont{Lust}},
  \bibinfo{journal}{Phys. Lett. B} \textbf{\bibinfo{volume}{855}},
  \bibinfo{pages}{138775} (\bibinfo{year}{2024}), \eprint{2312.12352}.

\bibitem[{\citenamefont{Toda et~al.}(2024)\citenamefont{Toda, Giar{\`e},
  {\"O}z{\"u}lker, Di~Valentino, and Vagnozzi}}]{Toda:2024ncp}
\bibinfo{author}{\bibfnamefont{Y.}~\bibnamefont{Toda}},
  \bibinfo{author}{\bibfnamefont{W.}~\bibnamefont{Giar{\`e}}},
  \bibinfo{author}{\bibfnamefont{E.}~\bibnamefont{{\"O}z{\"u}lker}},
  \bibinfo{author}{\bibfnamefont{E.}~\bibnamefont{Di~Valentino}},
  \bibnamefont{and} \bibinfo{author}{\bibfnamefont{S.}~\bibnamefont{Vagnozzi}},
  \bibinfo{journal}{Phys. Dark Univ.} \textbf{\bibinfo{volume}{46}},
  \bibinfo{pages}{101676} (\bibinfo{year}{2024}), \eprint{2407.01173}.

\bibitem[{\citenamefont{Akarsu et~al.}(2025)\citenamefont{Akarsu, Eingorn,
  Perivolaropoulos, Y{\"u}kselci, and Zhuk}}]{Akarsu:2025dmj}
\bibinfo{author}{\bibfnamefont{{\"O}.}~\bibnamefont{Akarsu}},
  \bibinfo{author}{\bibfnamefont{M.}~\bibnamefont{Eingorn}},
  \bibinfo{author}{\bibfnamefont{L.}~\bibnamefont{Perivolaropoulos}},
  \bibinfo{author}{\bibfnamefont{A.~E.} \bibnamefont{Y{\"u}kselci}},
  \bibnamefont{and} \bibinfo{author}{\bibfnamefont{A.}~\bibnamefont{Zhuk}}
  (\bibinfo{year}{2025}), \eprint{2504.07299}.

\bibitem[{\citenamefont{Di~Valentino et~al.}(2025)\citenamefont{Di~Valentino,
  Said, Riess, Pollo, Poulin, G{\'o}mez-Valent, Weltman, Palmese, Huang, van~de
  Bruck et~al.}}]{di2025cosmoverse}
\bibinfo{author}{\bibfnamefont{E.}~\bibnamefont{Di~Valentino}},
  \bibinfo{author}{\bibfnamefont{J.~L.} \bibnamefont{Said}},
  \bibinfo{author}{\bibfnamefont{A.}~\bibnamefont{Riess}},
  \bibinfo{author}{\bibfnamefont{A.}~\bibnamefont{Pollo}},
  \bibinfo{author}{\bibfnamefont{V.}~\bibnamefont{Poulin}},
  \bibinfo{author}{\bibfnamefont{A.}~\bibnamefont{G{\'o}mez-Valent}},
  \bibinfo{author}{\bibfnamefont{A.}~\bibnamefont{Weltman}},
  \bibinfo{author}{\bibfnamefont{A.}~\bibnamefont{Palmese}},
  \bibinfo{author}{\bibfnamefont{C.~D.} \bibnamefont{Huang}},
  \bibinfo{author}{\bibfnamefont{C.}~\bibnamefont{van~de Bruck}},
  \bibnamefont{et~al.}, \bibinfo{journal}{Physics of the Dark Universe}
  \textbf{\bibinfo{volume}{49}}, \bibinfo{pages}{101965}
  (\bibinfo{year}{2025}).

\bibitem[{\citenamefont{Akarsu et~al.}(2026)\citenamefont{Akarsu, Caruana,
  Dialektopoulos, Escamilla, Kahya, and Levi~Said}}]{Akarsu:2026anp}
\bibinfo{author}{\bibfnamefont{{\"O}.}~\bibnamefont{Akarsu}},
  \bibinfo{author}{\bibfnamefont{M.}~\bibnamefont{Caruana}},
  \bibinfo{author}{\bibfnamefont{K.~F.} \bibnamefont{Dialektopoulos}},
  \bibinfo{author}{\bibfnamefont{L.~A.} \bibnamefont{Escamilla}},
  \bibinfo{author}{\bibfnamefont{E.~O.} \bibnamefont{Kahya}}, \bibnamefont{and}
  \bibinfo{author}{\bibfnamefont{J.}~\bibnamefont{Levi~Said}}
  (\bibinfo{year}{2026}), \eprint{2602.08928}.

\bibitem[{\citenamefont{G{\"o}k{\c{c}}en
  et~al.}(2026)\citenamefont{G{\"o}k{\c{c}}en, Akarsu, and
  Di~Valentino}}]{Gokcen:2026pkq}
\bibinfo{author}{\bibfnamefont{M.}~\bibnamefont{G{\"o}k{\c{c}}en}},
  \bibinfo{author}{\bibfnamefont{{\"O}.}~\bibnamefont{Akarsu}},
  \bibnamefont{and}
  \bibinfo{author}{\bibfnamefont{E.}~\bibnamefont{Di~Valentino}}
  (\bibinfo{year}{2026}), \eprint{2602.21169}.

\bibitem[{\citenamefont{Basilakos et~al.}(2025)\citenamefont{Basilakos,
  Paliathanasis, and Saridakis}}]{Basilakos:2025olm}
\bibinfo{author}{\bibfnamefont{S.}~\bibnamefont{Basilakos}},
  \bibinfo{author}{\bibfnamefont{A.}~\bibnamefont{Paliathanasis}},
  \bibnamefont{and} \bibinfo{author}{\bibfnamefont{E.~N.}
  \bibnamefont{Saridakis}}, \bibinfo{journal}{Phys. Lett. B}
  \textbf{\bibinfo{volume}{868}}, \bibinfo{pages}{139658}
  (\bibinfo{year}{2025}), \eprint{2503.19864}.

\bibitem[{\citenamefont{Cai et~al.}(2010)\citenamefont{Cai, Saridakis, Setare,
  and Xia}}]{Cai:2009zp}
\bibinfo{author}{\bibfnamefont{Y.-F.} \bibnamefont{Cai}},
  \bibinfo{author}{\bibfnamefont{E.~N.} \bibnamefont{Saridakis}},
  \bibinfo{author}{\bibfnamefont{M.~R.} \bibnamefont{Setare}},
  \bibnamefont{and} \bibinfo{author}{\bibfnamefont{J.-Q.} \bibnamefont{Xia}},
  \bibinfo{journal}{Phys. Rept.} \textbf{\bibinfo{volume}{493}},
  \bibinfo{pages}{1} (\bibinfo{year}{2010}), \eprint{0909.2776}.

\end{thebibliography}

\end{document}